\documentclass[twocolumn]{aastex62}

\usepackage{amsmath}
\usepackage{lineno}
\received{January 1, 2018}
\revised{January 7, 2018}
\accepted{\today}
\submitjournal{ApJ}

\shorttitle{M dwarf's atmosphere \& wind}
\shortauthors{Sakaue \& Shibata}

\begin{document}

\title{M-dwarf's Chromosphere, Corona and Wind Connection via the Nonlinear Alfv\'en Wave}

\correspondingauthor{Takahito Sakaue}
\email{sakaue@kwasan.kyoto-u.ac.jp}


\author{Takahito Sakaue}
\affiliation{Astronomical Observatory, Kyoto University, Japan}

\author{Kazunari Shibata}
\affiliation{Astronomical Observatory, Kyoto University, Japan}
\affiliation{Doshisha University, Japan}






\begin{abstract}
  M dwarf's atmosphere is expected to be highly magnetized. The magnetic energy can be responsible for heating the stellar chromosphere and corona, and driving the stellar wind.
  The nonlinear propagation of Alfv\'en wave is the promising mechanism for both heating stellar atmosphere and driving stellar wind. Based on this Alfv\'en wave scenario, we carried out the one-dimensional compressive magnetohydrodynamic (MHD) simulation to reproduce the stellar atmospheres and winds of TRAPPIST-1, Proxima Centauri, YZ CMi, AD Leo, AX Mic, as well as the Sun. The nonlinear propagation of Alfv\'en wave from the stellar photosphere to chromosphere, corona, and interplanetary space is directly resolved in our study.
  The simulation result particularly shows that the slow shock generated through the nonlinear mode coupling of Alfv\'en wave is crucially involved in both dynamics of stellar chromosphere (stellar spicule) and stellar wind acceleration.
  Our parameter survey further revealed the following general trends of physical quantities of stellar atmosphere and wind.
  (1) The M dwarfs' coronae tend to be cooler and denser than solar corona.
  (2) M dwarfs' stellar winds can be characterized with relatively faster velocity and much smaller mass-loss rate compared to those of solar wind. The physical mechanisms behind these tendencies are clarified in this paper, where the stronger stratification of M dwarf's atmosphere and relatively smaller Alfv\'en wave energy input from the M dwarf's photosphere are remarkable.
\end{abstract}



\section{Introduction} \label{sec:intro}

M-type main sequence stars (M dwarfs) have the highly magnetized atmosphere. The magnetic energy generated in their convection zone emerges into the outer layer and contributes to heating the chromosphere and corona. The high energy radiation from these hot plasma is the manifestation of the stellar magnetic activity, and is observed in X-ray \citep{1996ApJ...463..707G,2000ApJ...533..372F,2003ApJ...594..982F,2004A&A...427..667N,2014ApJ...785....9W}, ultraviolet (UV) \citep{2013MNRAS.431.2063S,2016ApJ...820...89F}, optical lines such as H$\alpha$ and Ca II \citep{2000AJ....120.1085G,2009AJ....137.3297W}, and radio band \citep{1993ApJ...405L..63G,2010ApJ...709..332B}.
The multi-wavelength and spectroscopic observations have paid particular attention to the M-dwarf's flare and subsequent dynamics of stellar atmosphere \citep{1990A&A...238..249H,2018PASJ...70...62H,2019A&A...623A..49V,2020PASJ...72...68N,2021PASJ...73...44M}.

M dwarf's magnetic activities have been particularly discussed with the focus on their impact on the planetary atmosphere. The planets orbiting M dwarfs are favorable targets for the extrasolar habitable worlds \citep{1993Icar..101..108K,2007AsBio...7...30T,2007AsBio...7...85S,2009ApJ...698..519K,2013Sci...340..577S,2017ApJ...845....5K}. Their upper atmospheres are exposed to the high energy radiation in UV to X-ray range from the stellar atmosphere \citep{2009ApJ...703..905T,2012EP&S...64..179L,2015NatGe...8..177T,2016MNRAS.459.4088O} and affected by the stellar wind \citep{2011MNRAS.412..351V,2014MNRAS.438.1162V,2014ApJ...790...57C,2015ApJ...806...41C,2016ApJ...833L...4G,2017ApJ...843L..33G,2017ApJ...837L..26D,2018PNAS..115..260D,2020ApJ...902L...9A}. The resultant mass-loss from the planet's atmosphere determines its evolution especially for lower-mass planets.


Therefore, it is important for studies about the exoplanets or astrobiology to realize the underlying physics for the structure of stellar atmosphere and wind. Several numerical magnetohydrodynamics (MHD) modelings have been employed to explore the interplanetary environment around an M dwarf, but partly due to the lack of observational constraints, the theoretical predictions is not well established. 
For instance, the mass-loss rate of the stellar wind from TRAPPIST-1 (M8) is estimated to $\sim4.1\times10^{-15}M_\odot$ yr$^{-1}$ by the global three-dimensional (3D) MHD simulation of \citet{2018PNAS..115..260D} but to $3\times10^{-14}M_\odot$ yr$^{-1}$ by \citet{2017ApJ...843L..33G}. The simulated mass-loss rate of EV Lac (M3.5) by \citet{2014ApJ...790...57C} ($3\times10^{-14}M_\odot$ yr$^{-1}$) is four orders of magnitude higher than estimated by \citet{2011ApJ...741...54C}. Because these stellar wind modelings are sensitive to the inner boundary condition which represents the energy injection from the star to the interplanetary space \citep{2020MNRAS.494.1297M,2020A&A...635A.178B}, it is required to consider the connection between the stellar atmosphere and wind in more self-consistent manner. The lower atmosphere of M dwarf is characterized by the lower temperature, stronger stratification, higher density, smaller convective motion, and stronger magnetic field compared to the Sun \citep{2005nlds.book.....R}. Therefore, in order to discuss the diversity and universality of the stellar atmosphere and wind, it is inevitable to consider such properties unique to M dwarfs' lower atmosphere.

In addition to the connection between the stellar atmosphere and wind, the dynamics related to the nonlinear Alfv\'en wave is another important ingredient for the modeling of stellar atmosphere and wind \citep{1986JGR....91.4111H,1993A&A...270..304V}. The nonlinear propagation of Alfv\'en wave is the promising mechanisms for both heating the stellar atmosphere and driving the stellar wind. Alfv\'en wave transfers the magnetic energy efficiently in the magnetized plasma. The various nonlinear processes of Alfv\'en wave are responsible for the energy conversion from the magnetic energy to the kinetic or thermal energy of the background media \citep{1947MNRAS.107..211A,1961ApJ...134..347O,1968ApJ...153..371C,1983A&A...117..220H,1976ApJ...210..498B,1980JGR....85.1311H}.
  Owing to the high-resolution MHD simulations, it is found that, while the atmosphere and wind are maintained by the energy and momentum transfer by Alfvén waves, its propagation is affected by the dynamics of atmosphere such as spicule \citep{1982SoPh...75...35H,1999ApJ...514..493K,2010ApJ...710.1857M} and stellar wind \citep{2005ApJ...632L..49S,2012ApJ...749....8M,2013PASJ...65...98S,2014MNRAS.440..971M,2021MNRAS.500.4779M,2018PASJ...70...34S,2018ApJ...853..190S,2019ApJ...880L...2S,2020ApJ...896..123S}.

  These studies highlight the importance of resolving the relatively small-scale dynamics associated with the Alfv\'en wave propagation, as well as reproducing the global structure of the stellar atmosphere and wind.

In this study, therefore, we extend our recent solar atmosphere and wind model \citep{2020ApJ...900..120S} to the M dwarfs' atmosphere and wind. By carrying out the one-dimensional (1D) time-dependent MHD simulations, the nonlinear propagation of Alfv\'en wave in the nonsteady stellar atmosphere and wind is calculated from the M dwarf's photosphere to the chromosphere, corona, and interplanetary space. Part of the simulation results in this paper is also discussed in our previous paper \citep{2021ApJ...906L..13S}, in which we briefly summarized the similarities and differences in the structures of reproduced stellar atmosphere and wind among the Sun and M dwarfs. The present paper mainly focuses on the development of our semi-empirical method to estimate the stellar atmosphere and wind parameters (e.g., coronal temperature, wind velocity and mass-loss rate) based on the simulation results.

\begin{deluxetable*}{lcccccc}[t!]
\tablecaption{Parameters of stars. The effective temperature ($T_{\rm eff}$), surface gravity ($\log_{10}g$), stellar radius ($r_\star/r_\odot$), stellar mass ($M_\star/M_\odot$), pressure scale height of the photosphere ($H_{\rm ph}$), surface escape velocity ($v_{\rm esc\star}$). The physical quantities of photosphere, including  mass density ($\rho_{\rm ph}$), Rosseland opacity ($\kappa_R$), mean molecular weight ($\mu_{\rm ph}$), magnetic field strength ($B_{\rm ph}$), mixing length parameter ($\alpha_{\rm MLT}$), convective velocity ($v_{\rm conv}$), acoustic cutoff frequency ($\nu_{\rm ac}$). \label{table:table_stars}}
\tablecolumns{13}
\tablewidth{0pt}
\tablehead{
\colhead{Spectral Type} &
\colhead{G2} &
\colhead{M0} &
\colhead{M3.5} &
\colhead{M5} &
\colhead{M5.5} &
\colhead{M8}
}
\startdata
$T_{\rm eff}$ [K] & 5770  & 3800$^a$  & 3473$^c$ & 3280$^d$ & 3042$^f$ & 2559$^g$ \\
$\log_{10} g$ [cm s$^{-2}$] & 4.44 & 4.77$^b$ & 4.79$^c$ & 4.91$^e$ & 5.21$^f$ & 5.21$^g$ \\
$r_\star/r_\odot$ & 1 & 0.51$^b$ & 0.46$^c$ & 0.32$^e$ & 0.15$^f$ & 0.12$^g$ \\
$M_\star/M_\odot$ & 1 & 0.60$^b$ & 0.47$^c$ & 0.31$^e$ & 0.12$^f$ & 0.08$^g$ \\
$H_{\rm ph}$ [km] & 134 & 35.5 & 29.4 & 20.8 & 9.44 & 6.93 \\
$v_{\rm esc\star}$ [km s$^{-1}$] & 618 & 647 & 624 & 602 & 569 & 511 \\
$\rho_{\rm ph}$ [$\times10^{-7}$ g cm$^{-3}$] & 2.6 & 14 & 19 & 27 & 69 & 190 \\
$\kappa_R$ [cm$^2$ g$^{-1}$] & 0.19 & 0.14 & 0.13 & 0.11 & 0.10 & 0.053 \\
$\mu_{\rm ph}$ [g mol$^{-1}$] & 1.3 & 1.5 & 1.6 & 1.6 & 1.7 & 1.9 \\
$B_{\rm ph}$ [G] & 1560 & 2722 & 2936 & 3395 & 5139 & 7313 \\
$\alpha_{\rm MLT}$ & 1.6 & 2.1 & 2.4 & 2.6 & 3.1 & 4.0 \\
$v_{\rm conv}$ [km s$^{-1}$] ($v_{\rm conv}/c_{s,\rm ph}$) & 1.7 (0.21) & 0.59 (0.10) & 0.50 (0.09) & 0.42 (0.08) & 0.29 (0.06) & 0.18 (0.04) \\
$\nu_{\rm ac}$ [mHz] & 4.7 & 13 & 15 & 20 & 42 & 50 \\
typical star & Sun & AX Mic & AD Leo & YZ CMi & Proxima Centauri & TRAPPIST 1 \\
\enddata
\tablerefs{$^a$ \cite{2007ApJ...667..527G}; $^b$ \cite{2007ApJS..168..297T}; $^c$ \cite{2015AA...577A.132M}; $^d$ \cite{2014ApJ...791...54G}; $^e$ \cite{2017ApJ...834...85N}; $^f$ \cite{2003AA...397L...5S}; $^g$ \cite{2016Natur.533..221G}.}
\end{deluxetable*}

\section{NUMERICAL SETTING}
\subsection{Basic Equations}
\label{sec:basic_equations}

\begin{figure}
  \begin{center}
    \epsscale{1}
    \plotone{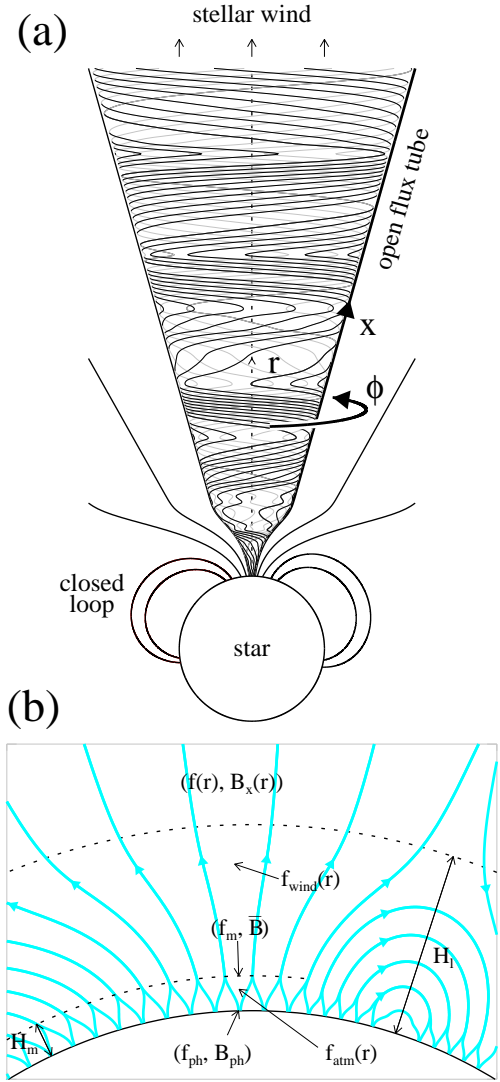}
    \caption{(a) Schematic drawing of axisymmetric magnetic flux tube, the surface of which is defined by the poloidal $x$ and toroidal $\phi$ axes. The winding thin lines represent the magnetic field lines, which illustrate the nonlinear propagation of Alfv\'en wave. (B) Schematic drawing of magnetic field configuration in the vicinity of stellar surface. The parameters defining the flux tube geometry in this study, such as $f_{\rm ph}$ and $H_m$, are indicated.}
    \label{fig:figure01.eps}
  \end{center}
\end{figure}

The non-linear propagation of the Alfv\'en wave in the time-dependent stellar atmosphere and wind is simulated by using 1D magnetohydrodynamic equations based on the axial symmetry assumption of the magnetic flux tube.
The surface of the axisymmetric flux tube is defined by the poloidal and toroidal axes which are noted in this study with $x$ and $\phi$ (Figure \ref{fig:figure01.eps}(a)). The basic equations are written as follows:

\begin{equation}
  {\partial\rho\over\partial t}+{1\over A}{\partial\over\partial x}(\rho v_xA)=0\label{eq:mass}
\end{equation}
\begin{align}
  {\partial\over\partial t}&\left({p\over\gamma-1}+{1\over2}\rho v^2+{B^2\over8\pi}\right)\nonumber\\
  +&{1\over A}{\partial\over\partial x}\left[A\left\{\left({\gamma p\over\gamma-1}+{\rho v^2\over2}+{B_\phi^2\over4\pi}\right)v_x-{B_x\over4\pi}(B_\phi v_\phi)\right\}\right]\nonumber\\
  &=\rho v_x{\partial\over\partial x}\left({GM_\star\over r}\right)-{1\over A}{\partial\over\partial x}(AF_c)-Q_{\mbox{\scriptsize rad}}\label{eq:ene}
\end{align}
\begin{align}
  {\partial(\rho v_x)\over\partial t}+&{\partial p\over\partial x}+{1\over A}{\partial\over\partial x}\left\{\left(\rho v_x^2+{B_\phi^2\over8\pi}\right)A\right\}\nonumber\\
  &-\rho v_\phi^2{\partial\ln\sqrt{A}\over\partial x}-\rho{\partial\over\partial x}\left({GM_\star\over r}\right)=0\label{eq:mom_pol}
\end{align}
\begin{equation}
  {\partial(\rho v_\phi)\over\partial t}+{1\over A\sqrt{A}}
  {\partial\over\partial x}\left\{A\sqrt{A}\left(\rho v_xv_\phi-{B_xB_\phi\over4\pi}\right)\right\}=0\label{eq:mom_tor}
\end{equation}
\begin{equation}
  {\partial B_\phi\over\partial t}+{1\over \sqrt{A}}{\partial\over\partial x}\Big(\sqrt{A}(v_xB_\phi-v_\phi B_x)\Big)=0\label{eq:mag_tor}
\end{equation}
\begin{equation}
  B_xA=\mbox{const.}\label{eq:mag_pol}
\end{equation}
\begin{equation}
  {dx\over dr}=\sqrt{1+\left({d\sqrt{A}\over dr}\right)^2}
\end{equation}
$\gamma$ represents the specific heat ratio and is set to 5/3 in this study. $F_c$ and $Q_{\rm rad}$ are the heat conduction flux and radiative cooling term, respectively, as described in Section \ref{sec:Heat Conduction and Radiative Cooling}. $r$ is the distance from the center of the Sun. $A$ is the cross section of the flux tube and is related to $r$ through the filling factor $f$ as $A(r)=4\pi r^2f(r)$. $f$ determines the geometry of the flux tube, which stems from the magnetic field concentration on the photosphere and expands to the interplanetary space.\par

\subsection{Magnetic Flux Tube Model}
\label{sec:magnetic_flux_tube_model}

The magnetic field strength of the photosphere $B_{\rm ph}$ is assumed to be equipartition to the ambient plasma pressure $p_{\rm ph}$. To estimate $p_{\rm ph}$, we refer to the opacity table presented by \cite{2014ApJS..214...25F}. The mass density of the photosphere $\rho_{\rm ph}$ would satisfy the following relation:
\begin{equation}
  \tau_R=\rho_{\rm ph}\kappa_R H_{\rm ph}=2/3
  \label{eq:determine_rho_ph}
\end{equation}
Here, $\tau_R$ and $\kappa_R$ are Rosseland optical depth and Rosseland opacity as a function of $(\rho,T)$. $\tau_R=2/3$ represents the effective depth of continuum formation \citep{1978stat.book.....M}. $H_{\rm ph}=R_gT_{\rm eff}/({\mu_{\rm ph}} g_\star)$ is the pressure scale height of the photosphere, where $R_g$ and $T_{\rm eff}$ are the gas constant and effective temperature, respectively. $\mu_{\rm ph}$ is the mean molecular weight of the photosphere, which is also presented by \cite{2014ApJS..214...25F}. By solving Equation (\ref{eq:determine_rho_ph}) about $\rho_{\rm ph}$ for given $T_{\rm eff}$ and $g_\star$, $p_{\rm ph}$ and $B_{\rm ph}$ can be determined by the equation of state $p_{\rm ph}=\rho_{\rm ph}R_gT_{\rm eff}/\mu_{\rm ph}$ and $B_{\rm ph}=\sqrt{8\pi p_{\rm ph}}$. These derived parameters of the stars are summarized in Table \ref{table:table_stars}.

Above the photosphere, magnetic flux tube expands exponentially so that the magnetic pressure inside the flux tube balances with the ambient plasma pressure decreasing with the scale height $H_{\rm ph}=R_gT_{\rm eff}/({\mu_{\rm ph}} g_\star)$. The filling factor $f$ in this layer is expected to be $f_{\rm atm}(r)=f_{\rm ph}\exp\{r_\star/(2H_{\rm ph})(1-r_\star/r)\}$, where $f_{\rm ph}$ is the filling factor of the photosphere. In the lower atmosphere where $r=r_\star+h$ ($h\ll r_\star$), we obtain $f_{\rm atm}(h)=f_{\rm ph}e^{h/(2H_{\rm ph})}$. This exponential expansion of flux tube stops at some height where it merges with the neighboring flux tube. Above this height, hereafter the merging height $H_m$, the magnetic pressure dominates the plasma pressure, and flux tube extends vertically with the constant cross section. The poloidal magnetic field strength is almost constant above $h=H_m$ through upper chromosphere and coronal base, and thus, $\overline{B}=B_{\rm ph}e^{-H_m/(2H_{\rm ph})}$ roughly represents the area-averaged magnetic field strength in the coronal hole from which the stellar wind emanates.

\begin{figure*}
  \begin{center}
    \epsscale{1} 
    \plotone{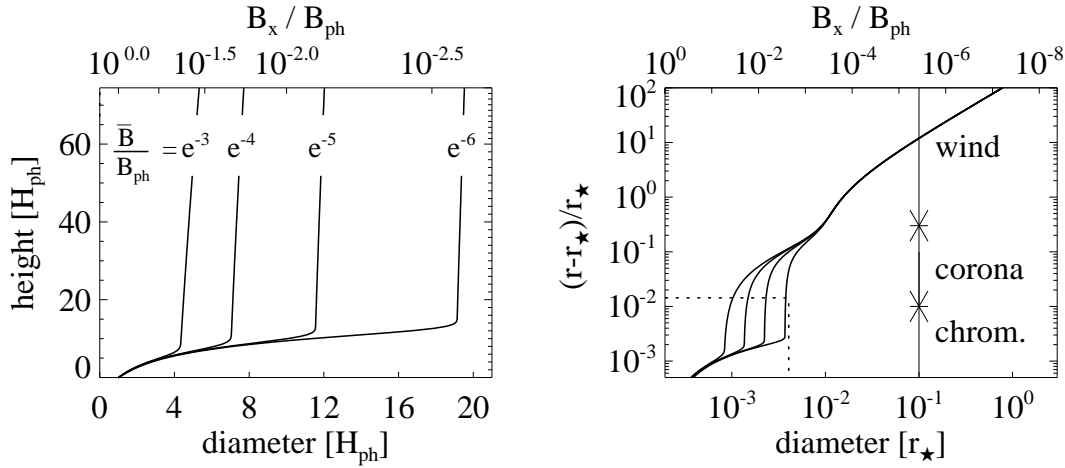}
    \caption{Poloidal magnetic field configurations characterized with the free parameter $\overline{B}$. Left and right panels show it in the lower and outer atmosphere. The dashed rectangle in the right panel corresponds to the ranges displayed in the left panel.}
    \label{fig:figure02.eps}
  \end{center}
\end{figure*}

The flux tube expands super radially again in the extended corona such that the interplanetary space is filled with the open flux tube. This expansion occurs around the height (i.e., loop height $H_l$), wherein the magnetic pressure of the closed loop significantly decreases. The functional form of the filling factor in this layer $f_{\rm wind}(r)$ is suggested by \citet{1976SoPh...49...43K}. Based on these considerations, the profile of the filling factor $f(r)$ is determined as follows:
\begin{equation}
  f_{\rm atm}(r)=f_m\tanh\left[{f_{\rm ph}\over f_m}\exp\left\{{r_\star\over 2H_{\rm ph}}\left(1-{r_\star\over r}\right)\right\}\right]
  \label{eq:f_eq1}
\end{equation}
where $f_m=f_{\rm ph}B_{\rm ph}/\overline{B}$
\begin{equation}
  f_{\rm wind}(r)={e^{(r-r_\star-H_l)/\sigma_l}+f_m-(1-f_m)e^{-(H_l/\sigma_l)}
    \over e^{(r-r_\star-H_l)/\sigma_l}+1}
  \label{eq:f_eq2}
\end{equation}
\begin{align}
  \hat{f}(r)=f_{\rm atm}(r)+{1\over2}&\Big(\max[f_{\rm wind}(r),f_m]-f_{\rm atm}(r)\Big)\nonumber\\
  &\times\left\{1+\tanh\left({r-r_\star-H_l\over H_l}\right)\right\}
  \label{eq:f_eq3}
\end{align}
\begin{equation}
  f(r)=f_{\rm ph}+(1-f_{\rm ph}){\hat{f}(r)-\hat{f}(r_\star)\over1-\hat{f}(r_\star)}
  \label{eq:f_eq4}
\end{equation}
$H_l$ and $\sigma_l$ in Eq. \ref{eq:f_eq2} are set to $0.1r_\star$.
The meanings of parameters used in the above definition are summarized in Figure \ref{fig:figure01.eps}(b).
We discuss the effect of varying $f_{\rm ph}$ in Appendix \ref{sec:appendix_fph}, and hereafter, focus on the simulation results with the fixed $f_{\rm ph}$ at 1/1600. When $f_{\rm ph}=1/1600$, the total open magnetic flux on the solar surface ($\Phi_{\rm open}=4\pi r_\star^2B_{\rm ph}f_{\rm ph}$) is $5.9\times10^{22}$ Mx, and $\Phi_{\rm open}/(10^{22}$ Mx) $=2.7,2.4,1.4,0.41,0.38$ for M0, M3.5, M5, M5.5 M8 dwarfs, respectively.
The configuration of the magnetic flux tube with $f_{\rm ph}$=1/1600 is depicted in Figure \ref{fig:figure02.eps}.
We carried out the parameter survey of $\overline{B}$, and used $\ln(\overline{B}/B_{\rm ph})=-3,-4,-5,-6$.

\subsection{Heat Conduction and Radiative Cooling}
\label{sec:Heat Conduction and Radiative Cooling}
The equation of state is $p=\rho R_gT/\mu(T)$. The radiative cooling $Q_{\rm rad}$ is considered as the empirical formulae, which is composed of three distinct terms; namely, the photospheric radiation $Q_{\rm ph}$, chromospheric  $Q_{\rm ch}$, and coronal radiation  $Q_{\rm co}$.
\begin{align}
  &\mu(T)=\left\{
    \begin{aligned}
      &\mu_{\rm ph\odot}\left(1-{\chi(T)\over2}\right) & & (T>T_{\rm eff\odot})\\
      &\mu_{\rm ph}+(\mu_{\rm ph\odot}-\mu_{\rm ph}){T-T_{\rm eff}\over T_{\rm eff\odot}-T_{\rm eff}} & & (T_{\rm eff}<T<T_{\rm eff\odot})
    \end{aligned}
    \right.
\end{align}
where $\chi(T)$ is the ionization degree as the function of temperature which calculated by referring to \citet{2012A&A...539A..39C}.
\begin{equation}
  Q_{\rm rad}=(1-\xi_1)(1-\xi_2)Q_{\rm ph}+\xi_1(1-\xi_2)Q_{\rm ch}+\xi_2Q_{\rm co}
\end{equation}
\begin{equation}
  \xi_1={1\over2}\left[1+\tanh\left({r-r_\star\over H_{\rm ph}}-3\right)\right]
\end{equation}
\begin{equation}
  \xi_2=\exp\left(-4\times10^{-20}\int_\infty^rn_{\rm HI}dr'\right)
\end{equation}
where $n_{\rm HI}$ is the neutral hydrogen number density; i.e., $n_{\rm HI}=(1-\chi(T))\rho/m_p$, where $m_p$ is the proton mass.
\begin{equation}
  Q_{\rm ph}=4\rho\kappa_R\sigma_{\rm SB} T^4\max\left({T^4\over T_{\rm ref}^4}-1,-e^{-(r-r_\star)^2/H_{\rm ph}^2}\right)
  \label{eq:eq_q_ph}
\end{equation}
\begin{equation}
  \mbox{where}\ \ T_{\rm ref}=T_{\rm eff}\left({3\over4}\rho\kappa_RH_{\rm ph}+{1\over2}\right)^{1/4}
\end{equation}
\begin{equation}
  Q_{\rm ch}=4.9\times10^9\ \mbox{[erg g$^{-1}$ s$^{-1}$]}\times\rho,\ \ \ Q_{\rm co}=\chi(T)n^2\Lambda(T)
\end{equation}
$\sigma_{SB}$ is Stefan–Boltzmann constant. $n$ is the number density of neutral or ionized hydrogen; i.e., $n=\rho/m_p$. $\Lambda(T)$ is the radiative loss function for the optically thin plasma. $Q_{\rm ch}$ and $\Lambda(T)$ are the same function as used in \citet{1997ApJ...489..426H}, which are always positive. The negative $Q_{\rm ph}$ in Eq. \ref{eq:eq_q_ph} represents the radiative heating and it is allowed only where $e^{-(r-r_\star)^2/H_{\rm ph}^2}\sim1$.

The heat conductive flux is composed of the collisional and collisionless term:
\begin{equation}
  F_{\rm c}=-\kappa(T){\partial T\over\partial x}
\end{equation}
\begin{equation}
  \kappa(T)=q\kappa_{\rm coll}+(1-q)\kappa_{\rm sat}
\end{equation}
\begin{equation}
  q=\max(0,\min(1,1-0.5\kappa_{\rm coll}/\kappa_{\rm sat}))
\end{equation}
\begin{equation}
  \kappa_{\rm sat}={3\over2}pv_{e,\rm thr}{r\over T}
\end{equation}
$\kappa_{\rm coll}(T)$ is adopted from \citet{1980SoPh...68..351N}, which considers the effect of partial ionization in lower temperature, while it agrees with Spitzer-H\"arm heat conductivity $\kappa_0 T^{5/2}$ (\cite{1953PhRv...89..977S}; $\kappa_0=10^{-6}$ in CGS unit) when $T>10^6$ K. $\kappa_{\rm sat}$ represents the saturation of heat flux due to the collisionless effect \citep{1964ApJ...139...93P,2013ApJ...769L..22B}. $v_{e,\rm thr}$ is the thermal speed of the electron. The above expression of $\kappa_{\rm sat}$ means the transition of heat conductivity from $\kappa_{\rm coll}$ to $\kappa_{\rm sat}$ occurs around $r\sim\lambda_{e,\rm mfp}$ ($\lambda_{e,\rm mfp}$ is the electron mean free path), and the heat flux is limited to ${3\over2}\alpha pv_{e,\rm thr}$ in the distance where $T\sim r^{-\alpha}$ ($\alpha=0.2-0.4$ for faster wind than $500$ km s$^{-1}$; \citet{1989JGR....94.6893M}). Based on the above heat conductivity, the heat conduction is solved by super-time-stepping method \citep{2012MNRAS.422.2102M,2014JCoPh.257..594M}

\subsection{Initial and Boundary Condition}
\label{sec:Initial_and_Boundary_Condition}

We set the static atmosphere with the temperature of $10^4$ K as the initial state.
The toroidal velocity $v_\phi$ on the bottom boundary represents the convective motion on the stellar photosphere. We define it by the frequency-dependent fluctuation with the following power spectrum. $v_{\rm ph}$ is the free parameter corresponding to the amplitude of the convective velocity. The phase offsets of fluctuation are randomly assigned.

\begin{equation}
  v_{\rm ph}^2\propto\int^{\nu_{\rm max}}_{\nu_{\rm min}}\nu^{-1}d\nu
\end{equation}
where we assumed that $\nu_{\rm min,max}=\nu_{\rm min,max\odot}(\nu_{\rm ac}/\nu_{\rm ac\odot})$ (see Table \ref{table:table_stars} for $\nu_{\rm ac}$ of each star), and that $\nu_{\rm min\odot}^{-1}$ and $\nu_{\rm max\odot}^{-1}$ are set to 30 minutes and 20 seconds, respectively. The amplitude of fluctuation $v_{\rm ph}$ is a subject of survey in this study. Based on the convection theory \citep{1984A&A...136..338B,1986AdSpR...6h..39U,2013A&A...559A..40S}, we estimate the fiducial convective velocity $v_{\rm conv}$ on the stellar photosphere based on the following scaling :
\begin{equation}
  v_{\rm conv}^3\propto{\alpha_{\rm MLT}T^4_{\rm eff}\over\rho_{\rm ph}}
\end{equation}
$\alpha_{\rm MLT}$ is the mixing length parameter which we derived by referring to \cite{1999A&A...346..111L,2002A&A...395...99L} and \cite{2015A&A...573A..89M}. The determined $\alpha_{\rm MLT}$ and $v_{\rm conv}$ for each star are summarized in Table \ref{table:table_stars}. For M dwarfs' atmospheres and winds, we carried out the simulations with $v_{\rm ph}/v_{\rm conv}=1.0,1.4,2.0,3.0$ in each case of $\ln(\overline{B}/B_{\rm ph})=-4,-5,-6$, and $v_{\rm ph}/v_{\rm conv}=1$ in the case of $\ln(\overline{B}/B_{\rm ph})=-3$. These used values of $v_{\rm ph}/v_{\rm conv}$ for solar and M dwarfs' atmospheres and winds simulations are summarized in Table \ref{table:table_v_ph}.

\begin{table}
  \begin{center}
    \caption{$v_{\rm ph}/v_{\rm conv}$ used in our parameter survey}
    \label{table:table_v_ph}
    \begin{tabular}{lcc}
      \hline
      $\ln(\overline{B}/B_{\rm ph})$ & $-3$ & $-4$, $-5$, $-6$ \\
      \hline
      Sun & 1.0 & 0.33, 0.67, 1.0, 2.0\\
      M dwarfs & 1.0 & 1.0, 1.4, 2.0, 3.0 \\
      \hline
    \end{tabular}
  \end{center}
\end{table}

To excite the purely outward Alfv\'en wave on the bottom boundary, the toroidal magnetic field $B_\phi$ is determined by $B_\phi=-\sqrt{4\pi\rho} v_\phi$. That means Els\"asser variables ($z_{\rm out}=v_\phi-B_\phi/\sqrt{4\pi\rho}$, $z_{\rm in}=v_\phi+B_\phi/\sqrt{4\pi\rho}$) on the bottom boundary satisfy the condition that $z_{\rm out}=2v_\phi$ and $z_{\rm in}=0$. The longitudinal velocity component $v_x$ on the bottom boundary is the same fluctuation as $v_\phi$ of the photosphere.

The upper boundary is treated as the free boundary. It corresponds to 100 $r_\star$ and 19200 grids are placed nonuniformly in-between. The numerical scheme is based on the HLLD Riemann solver \citep{2005JCoPh.208..315M} with the second-order MUSCL interpolation and the third-order TVD Runge–Kutta method \citep{1988JCoPh..77..439S}.

\section{Results}
\label{sec:results}
\subsection{Solar and Stellar Winds}
\begin{figure*}
  \begin{center}
    \epsscale{.9}
    \plotone{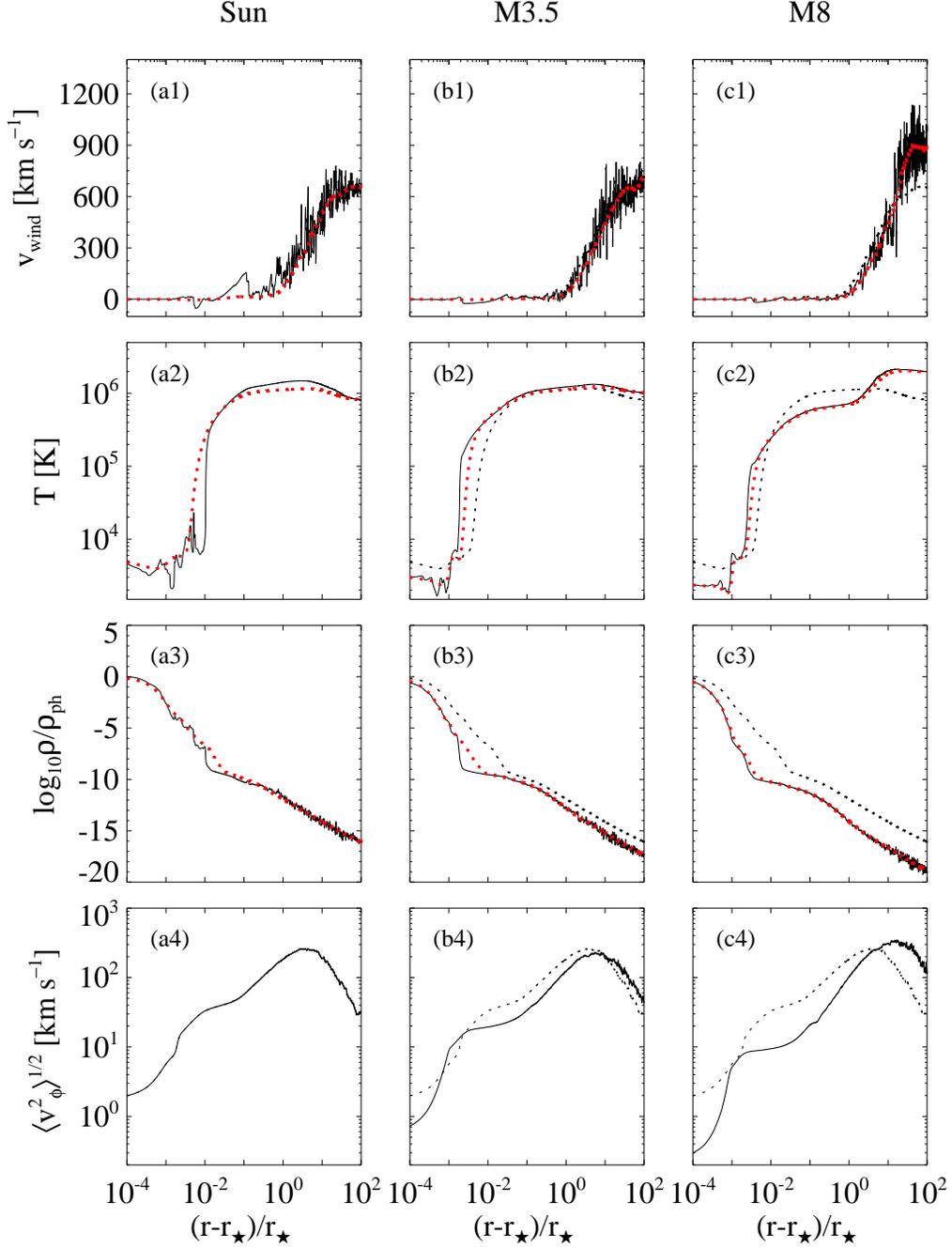}
    \caption{The snapshots of solar and stellar wind profiles about the velocity, temperature, mass density, and temporally averaged profiles of Alfv\'en wave amplitude. The left, middle and right columns correspond to the results of the solar wind, and stellar winds of M3.5 and M8 dwarfs, respectively. The results shown here are obtained by setting $v_{\rm ph}/v_{\rm conv}=1$ and $\ln(\overline{B}/B_{\rm ph})=-5$. The red dotted lines show the temporally averaged physical quantities of stellar wind, while the black dotted lines correspond to that of solar wind as a function of $(r-r_\star)/r_\star$.}
    \label{fig:figure03.eps}
  \end{center}
\end{figure*}

Figure \ref{fig:figure03.eps} shows the snapshots of solar and stellar wind profiles about the velocity, temperature, mass density, and temporally averaged profiles of Alfv\'en wave amplitude. The left, middle and right columns correspond to the results of the solar wind, and stellar winds of M3.5 and M8 dwarfs, respectively. The results shown here are obtained by setting $v_{\rm ph}/v_{\rm conv}=1$ and $\ln(\overline{B}/B_{\rm ph})=-5$. The red dotted lines show the temporally averaged physical quantities of stellar wind, while the black dotted lines correspond to that of solar wind as a function of $(r-r_\star)/r_\star$. As shown in Figure \ref{fig:figure03.eps}, the typical late M-dwarf's stellar wind is faster than the solar wind, and characterized with the smaller Alfv\'en wave amplitude in the lower corona $r<r_\star$.

\begin{figure*}
  \begin{center}
    \epsscale{1.2}
    \plotone{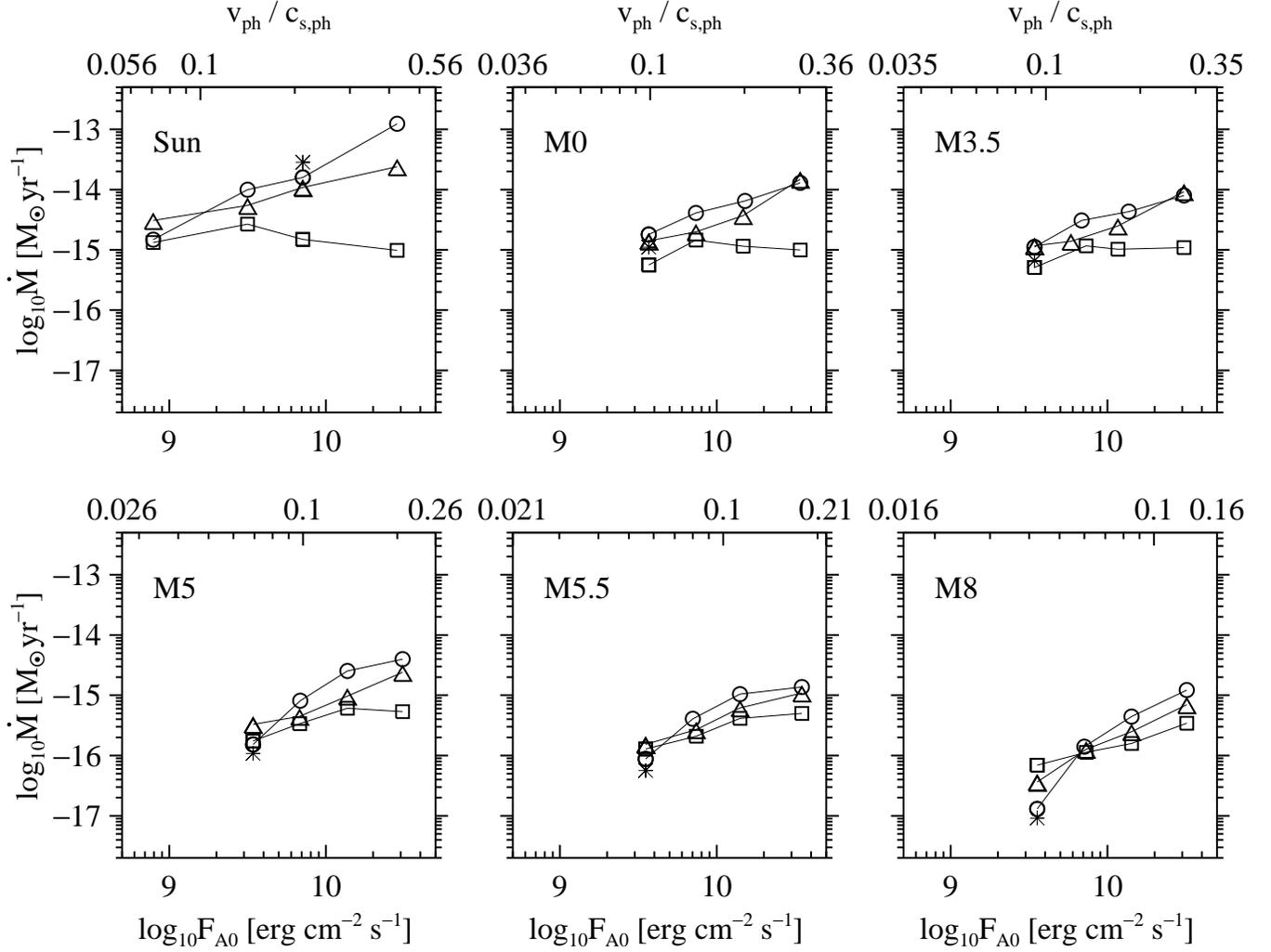}
    \caption{The mass-loss rates of solar and stellar winds as a function of the Alfv\'en wave energy flux of the photosphere ($F_{A0}=\rho_{\rm ph}v_{\rm ph}^2V_{A,\rm ph}$, where $V_{A,\rm ph}$ is the Alfv\'en speed of the photosphere).
      The upper horizontal axes represent the Mach number of velocity amplitude of the photosphere ($v_{\rm ph}/c_{s,\rm ph}$).
      The square, triangle, circle and asterisk symbols correspond to the results obtained by setting $\ln(\overline{B}/B_{\rm ph})=-6,-5,-4$ and $-3$, respectively. The fiducial values of $v_{\rm ph}/c_{s,\rm ph}$ are listed in Table \ref{table:table_stars}. In the case of the Sun, $v_{\rm ph}/c_{s,\rm ph}=0.21$ corresponds to $v_{\rm ph}/v_{\rm conv}=1$. Similarly, $v_{\rm ph}/c_{s,\rm ph}=$0.10, 0.09, 0.08, 0.06, 0.04 are expected as the fiducial values for M0, M3.5, M5, M5.5, M8 dwarfs, respectively.}
    \label{fig:figure04.eps}
  \end{center}
\end{figure*}

Figure \ref{fig:figure04.eps} shows the mass-loss rates of solar and stellar winds as a function of the Alfv\'en wave energy flux of the photosphere. The square, triangle, circle and asterisk symbols correspond to the results obtained by setting $\ln(\overline{B}/B_{\rm ph})=-6,-5,-4$ and $-3$, respectively. Table \ref{table:table_masslossrate} summarizes the mass-loss rates in the case of $v_{\rm ph}/v_{\rm conv}=1$ and shows that the wind's mass-loss rate of M dwarf is generally much smaller than the solar wind value. Another remarkable feature is that $\dot{M}$ as a function of $v_{\rm ph}$ is dependent on $\overline{B}/B_{\rm ph}$. When $\ln(\overline{B}/B_{\rm ph})=-6$ (square symbols), $\dot{M}$ of the solar wind and early M-dwarf's wind are less dependent on $v_{\rm ph}$. This phenomenon is well investigated by \cite{2020ApJ...900..120S}, who show that the magnetic energy cannot be transferred by Alfv\'en wave across the chromosphere when the nonlinearity of Alfv\'en wave is extremely high. 

\subsection{Solar and Stellar Spicules}
\label{sec:spicules}
\begin{figure*}
  \begin{center}
    \epsscale{1}
    \plotone{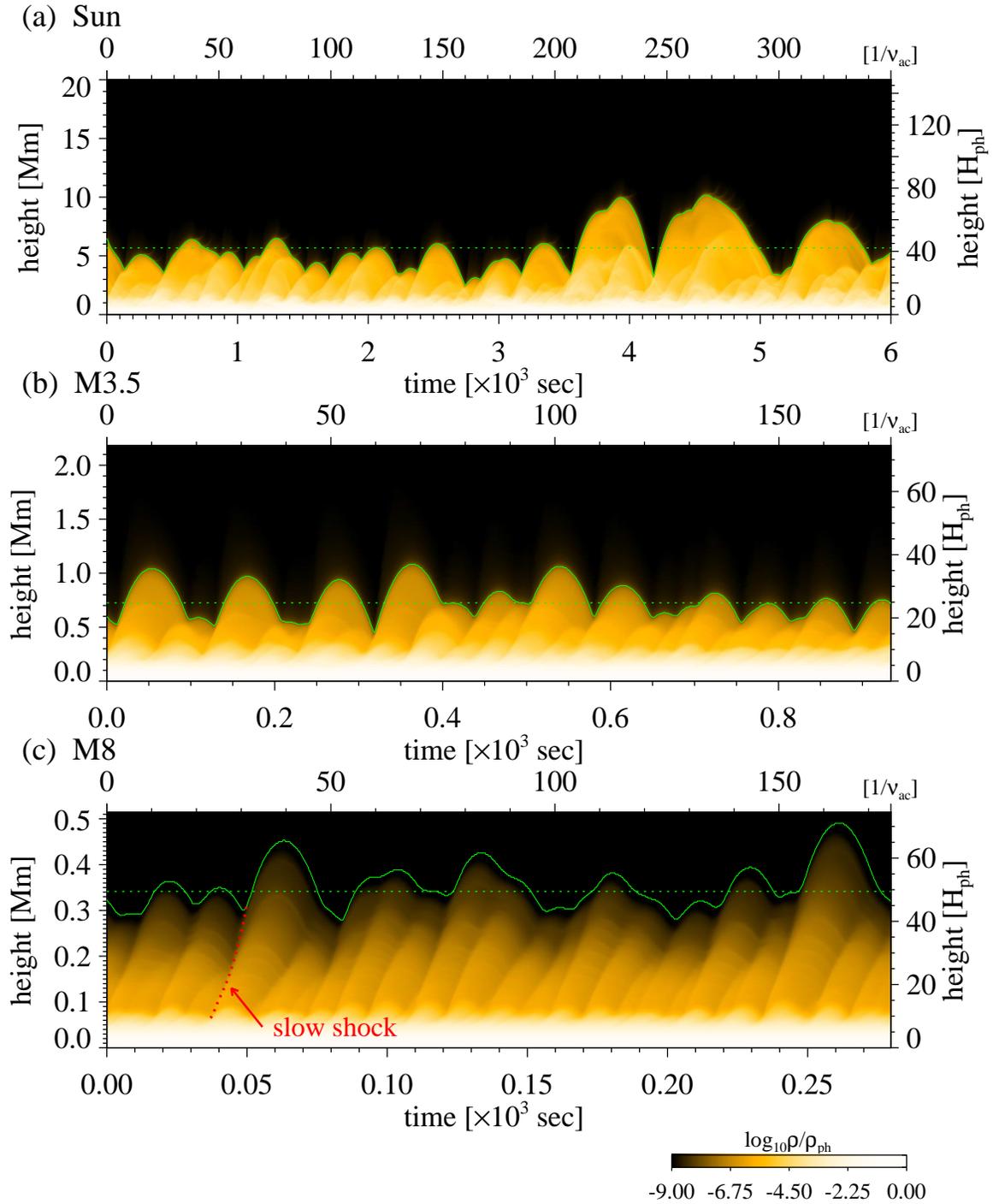}
    \caption{The time-slice diagrams of mass density in the lower atmospheres of (a) Sun, (b) M3.5 and (c) M8 dwarfs. The simulation results shown here are obtained by setting $v_{\rm ph}/v_{\rm conv}=1$ and $\ln(\overline{B}/B_{\rm ph})=-3$. The green solid lines in Figure \ref{fig:figure05.eps} represent the contour lines of $T=4\times10^4$ K (temperature of top of chromosphere). The upward and downward motions of top of chromosphere correspond to the dynamics of stellar spicule. The green horizontal dotted lines represent the median height of transition region ($H_{\rm tr}$).}
    \label{fig:figure05.eps}
  \end{center}
\end{figure*}

  Figure \ref{fig:figure05.eps} is the time-slice diagrams of mass density in the lower atmospheres of (a) Sun, (b) M3.5 and (c) M8 dwarfs. The simulation results shown here are obtained by setting $v_{\rm ph}/v_{\rm conv}=1$ and $\ln(\overline{B}/B_{\rm ph})=-3$. The green solid lines in Figure \ref{fig:figure05.eps} represent the contour lines of $T=4\times10^4$ K (temperature of top of chromosphere). The repetitive vertical motions of top of chromosphere correspond to the dynamics of stellar spicule.
  The onset of upward motion of spicule usually corresponds to the collision of chromospheric slow shock to the transition region. The trajectories of chromospheric slow shocks are clearly seen especially in panel (c) of Figure \ref{fig:figure05.eps}. An example of them are indicated by the red dotted line.

\begin{figure}
  \begin{center}
    \epsscale{1}
    \plotone{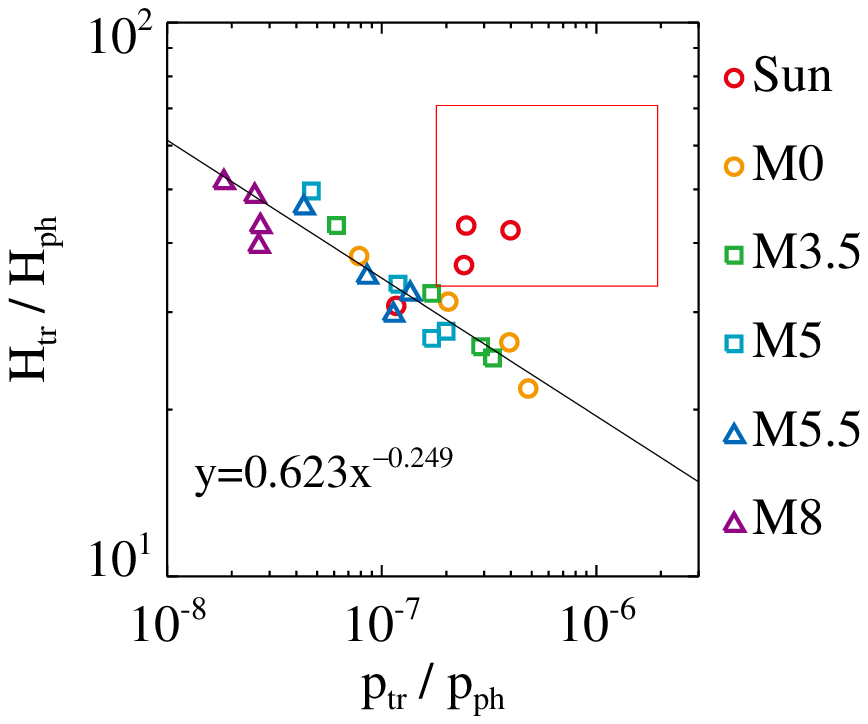}
    \caption{$H_{\rm tr}/H_{\rm ph}$ of various M dwarfs and Sun as a function of $p_{\rm tr}/p_{\rm ph}$. The simulation results shown here are obtained by setting $v_{\rm ph}/v_{\rm conv}=1$ and $\ln(\overline{B}/B_{\rm ph})=-3,-4,-5,-6$. The red rectangle represents the range of observed plasma pressure in solar spicule \citep{2018SoPh..293...20A,2020ApJ...888L..28S} and that of observed spicule height \citep{2012ApJ...750...16Z,2012ApJ...759...18P}.}
    \label{fig:figure06.eps}
  \end{center}
\end{figure}

  The green horizontal dotted lines represent the median height of transition region ($H_{\rm tr}$). It is notable that the normalized transition region height by pressure scale height of the photosphere ($H_{\rm tr}/H_{\rm ph}$) is the highest in M8 dwarf, while $H_{\rm tr}/H_{\rm ph}$ of M3.5 dwarf is smaller than that of the Sun. $H_{\rm tr}/H_{\rm ph}$ of various M dwarfs and the Sun are plotted in Figure \ref{fig:figure06.eps} as a function of $p_{\rm tr}/p_{\rm ph}$, where $p_{\rm tr}$ is the plasma pressure at the stellar transition region. The simulation results shown here are obtained by setting $v_{\rm ph}/v_{\rm conv}=1$ and $\ln(\overline{B}/B_{\rm ph})=-3,-4,-5,-6$.
  As shown in Figure \ref{fig:figure06.eps}, $H_{\rm tr}/H_{\rm ph}$ are well described by a negative power-law function of $p_{\rm tr}/p_{\rm ph}$ (black line), except for three cases corresponding to the solar transition region (red circles). These results lie within a red rectangle, which represents the range of observed plasma pressure in spicule \citep{2018SoPh..293...20A,2020ApJ...888L..28S} and that of observed spicule height \citep{2012ApJ...750...16Z,2012ApJ...759...18P}. Note that the observed spicule height means the maximum altitude that the transition region reaches, which is higher than the median height of transition region indicated by green dotted lines in Figure \ref{fig:figure05.eps}.

  Because higher transition region is associated with lower plasma pressure at transition region in the stratified atmosphere, the anticorrelation between $p_{\rm tr}/p_{\rm ph}$ and $H_{\rm tr}/H_{\rm ph}$ is naturally expected. The height of solar transition region is, on the other hand, significantly higher than the negative power-law function for M dwarfs'.
  This difference seems to be due to the stronger shocks in the solar chromosphere compared to those in M dwarfs' chromospheres. That means, even though $p_{\rm tr}/p_{\rm ph}$ of the Sun is comparable to that of M0 or M3.5 dwarfs, higher spicules can be driven by the stronger shocks in the solar chromosphere.
  As seen in Table \ref{table:table_stars}, the Mach number of convective motion of the photosphere ($v_{\rm conv}/c_{s,\rm ph}$) increases with the increasing effective temperature ($T_{\rm eff}$). The higher $v_{\rm conv}/c_{s,\rm ph}$, the higher nonlinearity of Alfv\'en wave in the chromosphere, which results in the stronger chromospheric shock.

\begin{table}
  \begin{center}
    \caption{$\dot{M}\ /(10^{-14}M_\odot\ {\rm yr}^{-1})$ ($v_{\rm ph}/v_{\rm conv}=1$)}
    \label{table:table_masslossrate}
    \begin{tabular}{lrrrr}
      \hline
      $\ln(\overline{B}/B_{\rm ph})$ &    $-3$ &    $-4$ &    $-5$ &    $-6$ \\
      \hline
      Sun &                  2.8 &                  1.6 &                  1.1 &   1.5$\times10^{-1}$ \\
      M0 &   1.1$\times10^{-1}$ &   1.8$\times10^{-1}$ &   1.4$\times10^{-1}$ &   5.6$\times10^{-2}$ \\
      M3.5 &   6.6$\times10^{-2}$ &   1.1$\times10^{-1}$ &   1.2$\times10^{-1}$ &   5.1$\times10^{-2}$ \\
      M5 &   1.1$\times10^{-2}$ &   1.5$\times10^{-2}$ &   3.3$\times10^{-2}$ &   1.7$\times10^{-2}$ \\
      M5.5 &   5.6$\times10^{-3}$ &   8.8$\times10^{-3}$ &   1.5$\times10^{-2}$ &   1.3$\times10^{-2}$ \\
      M8 &   9.2$\times10^{-4}$ &   1.3$\times10^{-3}$ &   3.6$\times10^{-3}$ &   6.9$\times10^{-3}$ \\
      \hline
    \end{tabular}
  \end{center}
\end{table}


\section{General Trends of Physical Quantities of Stellar Atmosphere and Wind}
\label{sec:analyses}

Numerical parameter survey reveals the general trends of various characteristics of stellar atmosphere and wind, including the wind velocity ($v_{\rm wind}$), mass-loss rate ($\dot{M}$), coronal temperature ($T_{\rm co}$), and plasma pressure at transition region ($p_{\rm tr}$). The differences and similarities in each of these parameters among the Sun and M dwarfs are described in this section. In the following, the coronal parameters are indicated with the subscript $_{\rm co}$ and represent those at $r=1.1r_\star$. The parameters at transition region are indicated with the subscript $_{\rm tr}$ as well.
The transition region is defined as the height with the temperature of $T_{\rm tr}=4\times10^4$ K. Because the transition region repeats the upward and downward motion violently (Figure \ref{fig:figure05.eps}), the physical quantities at the transition region should be defined as the temporally averaged values along the trajectory of the position of $T_{\rm tr}=4\times10^4$ K.

\subsection{Stellar Wind Acceleration}
\label{sec:vwind}
\begin{figure}
  \begin{center}
    \epsscale{1.2}
    \plotone{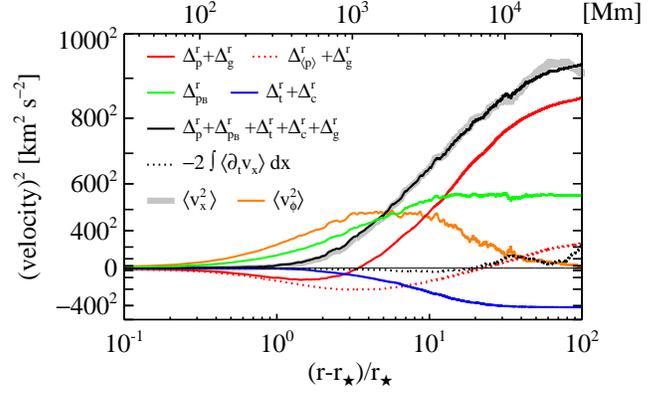}
    \caption{Bernoulli relation (Equation (\ref{eq:bernoulli})) about the squared stellar wind velocity (gray line) in the case of M3.5 dwarf with $v_{\rm ph}/v_{\rm conv}=1$ and $\ln(\overline{B}/B_{\rm ph})=-6$. $\Delta_p$ (Equation (\ref{eq:def_delta_p})), $\Delta_g$ (Equation (\ref{eq:def_delta_g})), $\Delta_{p_B}$ (Equation (\ref{eq:def_delta_p_B})), $\Delta_t$ (Equation (\ref{eq:def_delta_t})) and $\Delta_c$ (Equation (\ref{eq:def_delta_c})) represent the acceleration by the plasma pressure gradient, gravity, magnetic pressure gradient, magnetic tension force and centrifugal force. The red dotted line corresponds to the acceleration by the background plasma pressure gradient (Equation (\ref{eq:delta_p_back})).}
    \label{fig:figure07.eps}
  \end{center}
\end{figure}

To discuss the acceleration of the stellar wind, we pay attention to the poloidal component of equation of motion:

\begin{align}
  {\partial v_x\over\partial t}+&v_x{\partial v_x\over\partial x}+{1\over\rho}{\partial p\over\partial x}+{1\over\rho}{\partial\over\partial x}\left({B_\phi^2\over8\pi}\right)-v_\phi^2{\partial\ln\sqrt{A}\over\partial x}
  \nonumber\\
  &+{B_\phi^2\over4\pi\rho}{\partial\ln\sqrt{A}\over\partial x}-{\partial\over\partial x}\left(GM_\star\over r\right)=0
  \label{eq:eq_eom_vx}
\end{align}
After the temporal average and the integral from $r=r_\star$ to $r$ of both sides of Equation (\ref{eq:eq_eom_vx}), Bernoulli integral \citep{1963idp..book.....P,1999stma.book.....M} for the stellar wind velocity is obtained as follows:
\begin{align}
  v_x^2(r)=&-2\int^r_{r_\star}\left\langle{\partial v_x\over\partial t}\right\rangle dx\nonumber\\
  &+\Delta_p^r+\Delta_{p_B}^r+\Delta_c^r+\Delta_t^r+\Delta_g^r
  \label{eq:bernoulli}
\end{align}
where the first term of the right-hand side is negligible when the stellar wind is quasi-steady state. $\langle\cdot\rangle$ means the temporal average. The other terms are defined as follows:
\begin{equation}
  \Delta_p^r=-2\int^r_{r_\star}{1\over\rho}{\partial p\over\partial x}dx
  \label{eq:def_delta_p}
\end{equation}
\begin{equation}
  \Delta_{p_B}^r=-2\int^r_{r_\star}{1\over\rho}{\partial\over\partial x}\left({B_\phi^2\over8\pi}\right)dx
  \label{eq:def_delta_p_B}
\end{equation}
\begin{equation}
  \Delta_c^r=2\int^r_{r_\star} v_\phi^2{\partial\ln\sqrt{A}\over\partial x}dx
  \label{eq:def_delta_c}
\end{equation}
\begin{equation}
  \Delta_t^r=-2\int^r_{r_\star}{B_\phi^2\over4\pi\rho}{\partial\ln\sqrt{A}\over\partial x}dx
  \label{eq:def_delta_t}
\end{equation}
\begin{equation}
  \Delta_g^r=-v_{\rm esc\star}^2\left(1-{r_\star\over r}\right)
  \label{eq:def_delta_g}
\end{equation}
where $v_{\rm esc\star}=\sqrt{2GM_\star/r_\star}$. Equation (\ref{eq:bernoulli}) is confirmed in Figure \ref{fig:figure07.eps}, which shows the simulation result of stellar wind of M3.5 dwarf with $v_{\rm ph}/v_{\rm conv}=1$ and $\ln(\overline{B}/B_{\rm ph})=-6$. The black solid line corresponds to $\Delta_p^r+\Delta_g^r+\Delta_{p_B}^r+\Delta_t^r+\Delta_c^r$, which agrees well with $v_x^2$ (thick gray line) as indicated by Equation (\ref{eq:bernoulli}). It is most remarkable in Figure \ref{fig:figure07.eps} that the stellar wind is mainly accelerated by the plasma pressure gradient (red line). The magnetic pressure gradient (green line) contributes to supporting the stellar atmosphere and driving the stellar wind within $r\lesssim10r_\star$, but not involved in the further acceleration of stellar wind beyond the distance where the Alfv\'en wave amplitude (orange line) reaches a maximum. The magnetic tension force decelerates the stellar wind against the acceleration by the centrifugal force (blue line).

The above-mentioned acceleration by the plasma pressure gradient is distinguished from the hydrodynamic expansion proposed by \citet{1958ApJ...128..664P}. That means the acceleration by the plasma pressure gradient in our simulation is much larger than expected from the steady profile of the plasma pressure. By defining $\Delta_{\langle p\rangle}^r$ as below, we can see the acceleration by the background plasma pressure gradient $\Delta_{\langle p\rangle}^r+\Delta_g^r$ is much less effective to produce the simulated stellar wind velocity (red dotted line in Figure \ref{fig:figure07.eps}).
\begin{equation}
  \Delta_{\langle p\rangle}^r=-2\int^r_{r_\star}{1\over\langle\rho\rangle}{\partial\langle p\rangle\over\partial x}dx
  \label{eq:delta_p_back}
\end{equation}

\begin{figure}
  \begin{center}
    \epsscale{1.2}
    \plotone{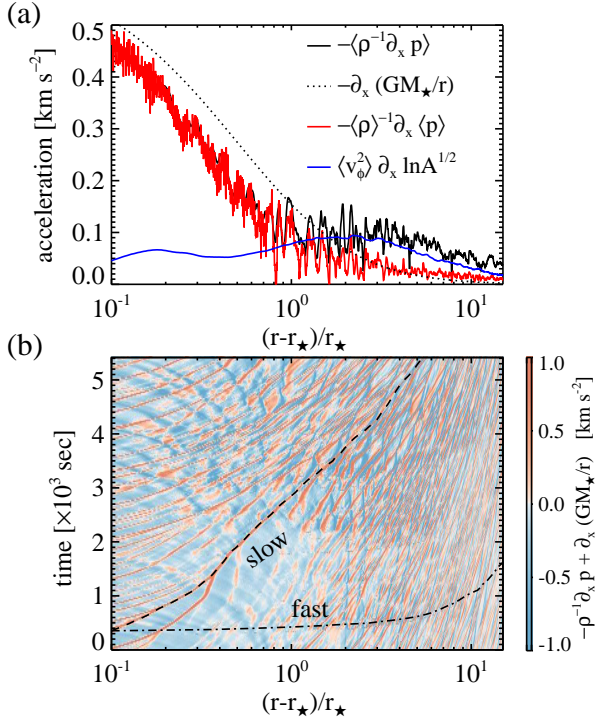}
    \caption{Panel (a): The comparison between the temporal average of the plasma pressure gradient $-\langle\rho^{-1}\partial_xp\rangle$ (black line) with the plasma pressure gradient calculated from the temporal average of the mass density and plasma pressure $-\langle\rho\rangle^{-1}\partial_x\langle p\rangle$ (red line). Panel (b): The time slice diagram of the plasma pressure gradient after subtracting the gravitational acceleration. The positive value (red) corresponds to the acceleration, while the negative value (blue) does to the deceleration. The dashed and dash-dotted lines correspond to the propagation of slow and fast mode waves, respectively.}
    \label{fig:figure08.eps}
  \end{center}
\end{figure}

Figure \ref{fig:figure08.eps}(a) shows the comparison between the temporal average of the plasma pressure gradient $-\langle\rho^{-1}\partial_xp\rangle$ (black line) with the plasma pressure gradient calculated from the temporal average of the mass density and plasma pressure $-\langle\rho\rangle^{-1}\partial_x\langle p\rangle$ (red line). Beyond $r\sim2r_\star$, we can see $-\langle\rho^{-1}\partial_xp\rangle$ surpasses $-\langle\rho\rangle^{-1}\partial_x\langle p\rangle$ and becomes comparable to the centrifugal force (blue line). Figure \ref{fig:figure08.eps}(b) is the time slice diagram of the plasma pressure gradient after subtracting the gravitational acceleration. The dashed and dash-dotted lines correspond to the typical characteristics of slow and fast mode waves, respectively. This figure shows the signatures with stronger plasma pressure gradient propagates at slow mode speed through $r\sim 2r_\star$ to $r\sim10r_\star$. Therefore, it is concluded that the numerous slow shocks excited around these distances lead to larger $-\langle\rho^{-1}\partial_xp\rangle$ than $-\langle\rho\rangle^{-1}\partial_x\langle p\rangle$, and significantly contribute to the stellar wind acceleration.

\begin{figure}
  \begin{center}
    \epsscale{1.}
    \plotone{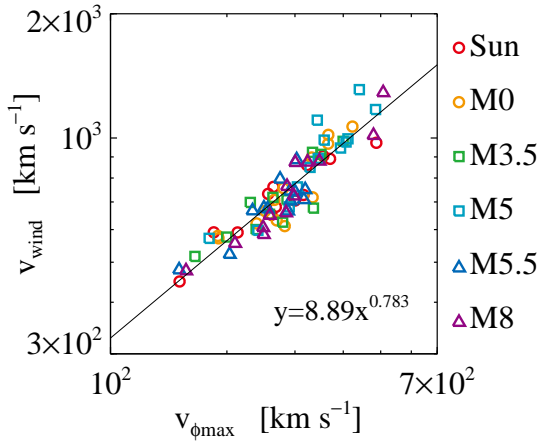}
    \caption{The wind velocity ($v_{\rm wind}$) as a function of maximum amplitude of Alfv\'en wave in the stellar wind ($v_{\phi\max}$). The solid line represents the power-law fitting.}
    \label{fig:figure09.eps}
  \end{center}
\end{figure}

Because the slow shock in stellar wind is excited by nonlinear Alfv\'en wave, faster wind velocity is expected when Alfv\'en wave is more amplified in the stellar wind. The resultant strong correlation between wind velocity ($v_{\rm wind}$) and maximum amplitude of Alfv\'en wave in the stellar wind ($v_{\phi\max}$) is shown in Figure \ref{fig:figure09.eps}.

\subsection{Coronal Temperature $T_{\rm co}$}
\label{sec:coronal_temperature}

The coronal temperature ($T_{\rm co}$) is determined by the balance between the heat conduction flux and transmitted Poynting flux into the corona, according to the following energy conservation law in the quasi-steady stellar wind.
\begin{equation}
  {\partial\over\partial x}[A(F_A+F(v_x)+F_g+F_c+F_{\rm rad})]=0,
  \label{eq:energy_conservation_der}
\end{equation}
where $F_A$, $F_g$, $F_c$ and $F_{\rm rad}$ are the Poynting flux carried by the Alfv\'en wave, gravitational energy flux, heat conduction flux, and radiative loss flux, respectively.
\begin{equation}
  F_A=-{B_x\langle B_\phi v_\phi\rangle\over4\pi},
\end{equation}
\begin{equation}
  F_g=-\langle\rho v_x\rangle {GM_\star\over r},
\end{equation}
\begin{equation}
  F_{\rm rad}={1\over A}\int^x_\infty A\langle Q_{\rm rad}\rangle dx.
\end{equation}

$F(v_x)$ is the sum of enthalpy flux $F_{\rm ent}$, kinetic energy flux $F_{\rm kin}$ and the Poynting flux advected with the stellar wind. Namely,
\begin{equation}
  F_{\rm ent}=\langle\gamma pv_x/(\gamma-1)\rangle,
\end{equation}
\begin{equation}
  F_{\rm kin}=\langle\rho v^2v_x/2\rangle,
\end{equation}
\begin{equation}
  F(v_x)=F_{\rm ent}+F_{\rm kin}+\langle B_\phi^2v_x/(4\pi)\rangle.
\end{equation}

\begin{figure}
  \begin{center}
    \epsscale{1.2}
    \plotone{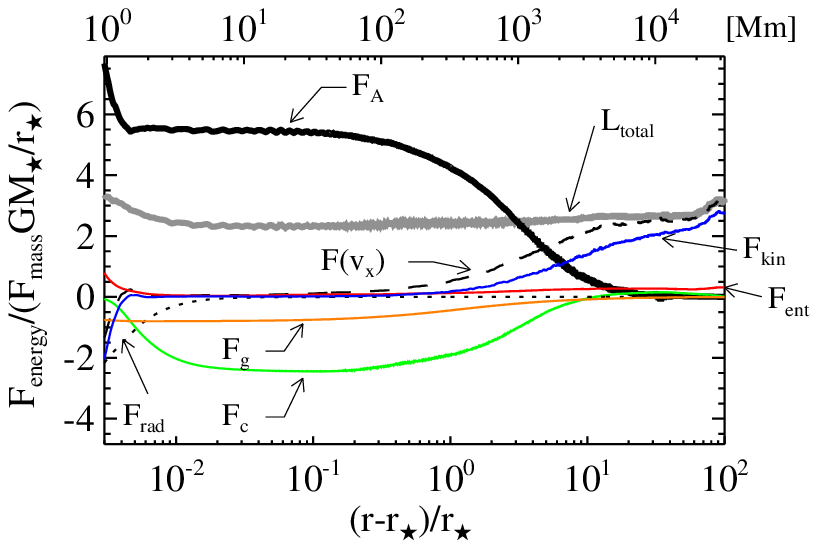}
    \caption{The energy flux conservation law in the quasi-steady stellar wind. The profile of each term in Equation (\ref{eq:energy_conservation}) in the case of M3.5 dwarf with $\ln(\overline{B}/B_{\rm ph})=-6$ and $v_{\rm ph}/v_{\rm conv}=1$ is shown. The energy fluxes are normalized by $F_{\rm mass}GM_\star/r_\star$, where $F_{\rm mass}$ is the mass flux and $F_{\rm mass}GM_\star/r_\star\approx5\times10^3r_\star^2/(fr^2)$ erg cm$^{-2}$ s$^{-1}$. $F_A$, $F_{\rm kin}$, $F_{\rm ent}$, $F_g$, $F_{\rm rad}$, $F_c$, and $F(v_x)$ are Alfv\'en wave energy flux, kinetic energy flux, enthalpy flux, gravitational energy flux, heat conduction flux, and the sum of enthalpy flux, kinetic energy flux, and Poynting flux advected with the stellar wind. $L_{\rm total}$ is the integral constant in Equation (\ref{eq:energy_conservation}).
    }
    \label{fig:figure10.eps}
  \end{center}
\end{figure}

Equation (\ref{eq:energy_conservation_der}) means that there is an integral constant $L_{\rm total}=F_{\rm total}A$ corresponding to the conservation of total energy flux.
\begin{equation}
  L_A+L(v_x)+L_g+L_c+L_{\rm rad}=L_{\rm total},
  \label{eq:energy_conservation}
\end{equation}
where $L_A=F_AA$, $L(v_x)=F(v_x)A$, $L_g=F_gA$, $L_c=F_cA$, and $L_{\rm rad}=F_{\rm rad}A$.
Figure \ref{fig:figure10.eps} shows the profile of each term in Equation (\ref{eq:energy_conservation}) in the case of M3.5 dwarf with $\ln(\overline{B}/B_{\rm ph})=-6$ and $v_{\rm ph}/v_{\rm conv}=1$. The energy fluxes are normalized by $F_{\rm mass}GM_\star/r_\star$, where $F_{\rm mass}$ is the mass flux and $F_{\rm mass}GM_\star/r_\star\approx5\times10^3r_\star^2/(fr^2)$ erg cm$^{-2}$ s$^{-1}$.

In our simulation, $L_{\rm total}$ is almost constant above $1.01r_\star$, below which the stellar atmosphere is too dynamic to rely on quasi-steady approximation. 
At the coronal height ($r-r_\star\sim0.1r_\star$), the energy balance is satisfied among $F_A$, $F_g$, and $F_c$, while in the distance ($\gtrsim 10r_\star$), the kinetic energy flux of the stellar wind ($F_{\rm kin}$) dominates the total energy flux. The radiative energy loss ($F_{\rm rad}$) is almost negligible at the coronal height in Figure \ref{fig:figure10.eps}. Therefore, by defining $L_{A,\rm co}$, $L_{g,\rm co}$, $L_{c,\rm co}$ as the energy luminosities $F_AA$, $F_gA$, $F_cA$ at $r=1.1r_\star$ and $L_{\rm kin,wind}$ as $F_{\rm kin}A$ at $r=100r_\star$, the energy conservation along the magnetic flux tube is approximately expressed as
\begin{equation}
  L_{A,\rm co}\approx L_{\rm kin,wind}-L_{g,\rm co}-L_{c,\rm co}.
  \label{eq:energy_conservation_app}
\end{equation}

The coronal temperature ($T_{\rm co}$) is related to the transmitted Alfv\'en wave energy flux ($L_{A,\rm co}$) through the above equation and the definition of Spitzer-H\"arm heat conductivity.
\begin{equation}
  \kappa_0T^{5/2}{\partial T\over\partial x}=-{L_c\over A}.
\end{equation}

By integrating the above from $x_{\rm tr}$ to $x_{\rm co}$ and neglecting $T_{\rm tr}/T_{\rm co}\ll 1$, we obtain:
\begin{equation}
  T_{\rm co}^{7/2}\approx-{7\over2\kappa_0}\int^{x_{\rm co}}_{x_{\rm tr}}dx{L_c\over A}.
\end{equation}
When we assume that $L_c$ is almost constant at $L_{c,\rm co}$ from $x_{\rm tr}$ to $x_{\rm co}$, as suggested by Figure \ref{fig:figure10.eps}, $T_{\rm co}$ is estimated as below.
\begin{equation}
  T_{\rm co}\approx\left[7|F_{c,\rm co}|l_{B,\rm co}\over 2\kappa_0\right]^{2/7},
  \label{eq:fcco_vs_tco}
\end{equation}
where $l_{B,\rm co}$ represents the spatial scale of expanding magnetic flux tube.
\begin{equation}
  l_{B,\rm co}=\int^{x_{\rm co}}_{x_{\rm tr}}dx{A_{\rm co}\over A}.
  \label{eq:l_b_definition}
\end{equation}
The similar scaling relation has been considered to discuss the temperature of quiescent or flaring coronal loop \citep{1978ApJ...220..643R,1998ApJ...494L.113Y}.

\begin{figure}
  \begin{center}
    \epsscale{1.}
    \plotone{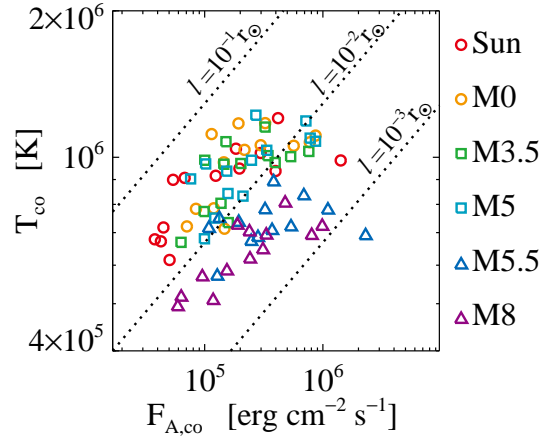}
    \caption{The coronal temperature ($T_{\rm co}$) as a function of transmitted Alfve\'en wave energy flux into the corona ($F_{A,\rm co}$). The dotted lines represent $y=\left[7xl/(2\kappa_0)\right]^{2/7}$ with $l=(10^{-1}$, $10^{-2}$, $10^{-3})r_\odot$.}
    \label{fig:figure11.eps}
  \end{center}
\end{figure}

Figure \ref{fig:figure11.eps} shows the relation between $F_{A,\rm co}$ and $T_{\rm co}$. As a result of tight correlation between $F_{A,\rm co}$ and $|F_{c,\rm co}|$, the power-law relation similar to Equation (\ref{eq:fcco_vs_tco}) can be seen in Figure \ref{fig:figure11.eps}. It is also shown that $T_{\rm co}$ of late M dwarfs (especially M5.5 and M8 dwarfs) is systematically lower than the solar coronal temperature with respect to a given $F_{A,\rm co}$. Equation (\ref{eq:fcco_vs_tco}) suggests that this difference originates in the much smaller spatial scale of magnetic flux tube ($l_{B,\rm co}$) of late M dwarfs, compared to the Sun. The dotted lines in Figure \ref{fig:figure11.eps} represent $y=\left[7xl/(2\kappa_0)\right]^{2/7}$ with $l=(10^{-1}$, $10^{-2}$, $10^{-3})r_\odot$.

\subsection{Plasma Pressure at Transition Region $p_{\rm tr}$}
\label{sec:p_tr}

\begin{figure}
  \begin{center}
    \epsscale{1.}
    \plotone{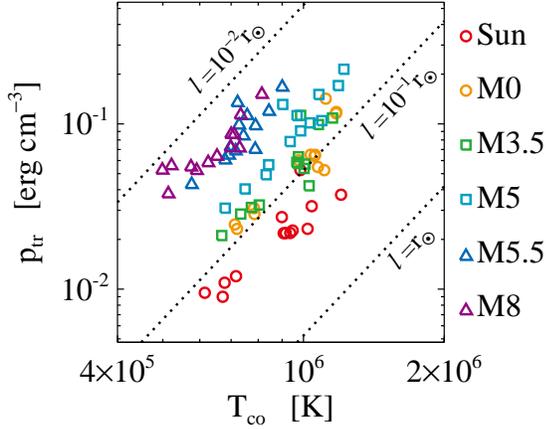}
    \caption{The plasma pressure at transition region ($p_{\rm tr}$) as a function of coronal temperature ($T_{\rm co}$). The dotted lines represent $y=\left(x/1.4\times10^3\right)^3/l$ with $l=(1$, $10^{-1}$, $10^{-2})r_\odot$.}
    \label{fig:figure12.eps}
  \end{center}
\end{figure}

The coronal temperature ($T_{\rm co}$) is one of the fundamental parameters determining the plasma pressure at transition region ($p_{\rm tr}$). \cite{1978ApJ...220..643R} pointed out that the energy balance along the coronal loop between the heat conduction flux and the radiative energy loss leads to the power-law relation among the coronal loop temperature ($T_{\rm loop}$), plasma pressure ($p_{\rm loop}$) and loop length ($l$); i.e., $T_{\rm loop}\approx1.4\times10^3(p_{\rm loop}l)^{1/3}$ in cgs unit (RTV scaling). Although their assumption of constant pressure along the coronal loop is not straightforwardly applicable to our case of open flux tube, the similar power-law relation between $T_{\rm co}$ and $p_{\rm tr}$ is seen in our simulation results. Figure \ref{fig:figure12.eps} shows the relation between $T_{\rm co}$ and $p_{\rm tr}$ obtained from our simulation. It is seen that $p_{\rm tr}$ of late M dwarfs (especially M5.5 and M8 dwarfs) is systematically higher than the solar value with respect to a given $T_{\rm co}$. The dotted lines represent RTV scaling; $y=\left(x/1.4\times10^3\right)^3/l$ with $l=(1$, $10^{-1}$, $10^{-2})r_\odot$.

\section{Semi-empirical Method to Predict the Characteristics of Stellar Atmosphere and Wind}
\label{sec:empirical}

The present study aims at establishing the empirical formulae to estimate the physical quantities of M dwarf's atmosphere and wind, including the wind velocity ($v_{\rm wind}$), mass-loss rate ($\dot{M}$), coronal temperature ($T_{\rm co}$), and plasma pressure at transition region ($p_{\rm tr}$).

\subsection{Stellar Wind Velocity vs Plasma $\beta$ of Stellar Wind}
\label{sec:vwind_empirical}

Based on the discussion in Section \ref{sec:vwind}, $v_{\rm wind}$ (wind velocity at $r=100r_\star$) is expressed as below.
\begin{equation}
  v_{\rm wind}^2=\Delta_p+\Delta _{p_B}+\Delta_c+\Delta_t+\Delta_g,
  \label{eq:vwind_estimate}
\end{equation}
where $\Delta_p=\Delta_p^{100r_\star}$, $\Delta_{p_B}=\Delta_{p_B}^{100r_\star}$, $\Delta_c=\Delta_c^{100r_\star}$, $\Delta_t=\Delta_t^{100r_\star}$, and $\Delta_g=\Delta_g^{100r_\star}$ (see Equations  (\ref{eq:def_delta_p})-(\ref{eq:def_delta_g})).

The maximum amplitude of Alfv\'en wave in the stellar wind ($v_{\phi\max}$) characterizes the above terms as follows:
\begin{equation}
  \Delta_p+\Delta_g=a_{1,1}\tilde{v}_{\phi\max}^{k_{1,1}},
\end{equation}
\begin{equation}
  \Delta_c=a_{1,2}\tilde{v}_{\phi\max}^{k_{1,2}},
\end{equation}
\begin{equation}
  \Delta_t=-a_{1,3}\tilde{v}_{\phi\max}^{k_{1,3}},
\end{equation}
\begin{equation}
  \Delta_{p_B}=a_{1,4}|\tilde{\Delta}_c+\tilde{\Delta}_t|^{k_{1,4}},
\end{equation}
where $\tilde{v}_{\phi\max}=v_{\phi\max}/($300 km s$^{-1}$), $\tilde{\Delta}_c=\Delta_c/(319^2$ (km s$^{-1})^2$), $\tilde{\Delta}_t=\Delta_t/(319^2$ (km s$^{-1})^2$). The coefficients ($a_{1,1}$, $a_{1,2}$, $a_{1,3}$, $a_{1,4}$) and power-law indices ($k_{1,1}$, $k_{1,2}$, $k_{1,3}$, $k_{1,4}$) are determined by regression analyses about our simulation results
\[
  a_{1,1}=653^2,\ \ \ \ \
  a_{1,2}=585^2,\ \ \ \ \
  a_{1,3}=666^2,\ \ \ \ \
  a_{1,4}=472^2,
\]
in unit of (km s$^{-1})^2$.
\[
  k_{1,1}=2.31,\ \ \ \ \
  k_{1,2}=2.04,\ \ \ \ \
  k_{1,3}=2.12,\ \ \ \ \
  k_{1,4}=0.682.
\]

\begin{figure}
  \begin{center}
    \epsscale{1.}
    \plotone{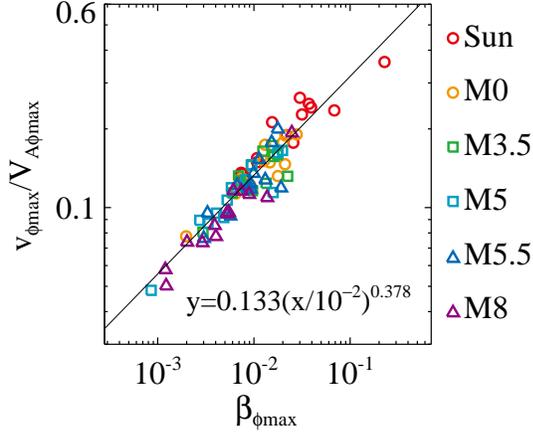}
    \caption{The maximum nonlinearity of Alfv\'en wave in the stellar wind vs plasma $\beta$ at the distance where the amplitude of Alfv\'en wave reaches the maximum.}
    \label{fig:figure13.eps}
  \end{center}
\end{figure}

As for $v_{\phi\max}$, we pay attention to the relation between $v_{\phi\max}$ and plasma $\beta$ in the stellar wind. Our simulation results clearly show the following positive correlation between $v_{\phi\max}/V_{A\phi\max}$ and $\beta_{\phi\max}$ (Figure \ref{fig:figure13.eps}), where $V_{A\phi\max}$ and $\beta_{\phi\max}$ are Alfv\'en speed and plasma $\beta$ at the distance where Alfv\'en wave amplitude reaches a maximum.
\begin{equation}
  {v_{\phi\max}\over V_{A\phi\max}}=0.133\left(\beta_{\phi\max}\over10^{-2}\right)^{0.378}.
  \label{eq:eq_nonlinearity_beta}
\end{equation}

The positive correlation between the nonlinearity of Alfv\'en wave and plasma $\beta$ of background media has been discussed in terms of decay instability \citep{1969npt..book.....S,1978ApJ...224.1013D}.
It is well known that the circularly polarized Alfv\'en wave with the frequency $\nu$ is susceptible to the decay instability in the low $\beta$ plasma, whose growth rate $\gamma$ is approximately expressed with $\gamma^2/\nu^2={1\over4}\eta^2\beta^{-1/2}$. $\eta=v_\perp/V_A$ is the nonlinearity of circularly polarized Alfv\'en wave with the amplitude of $v_\perp$. This suggests that the Alfv\'en wave propagation is drastically destabilized within the timescale comparable to its wave period, when $\eta\gtrsim 2\beta^{1/4}$. Although the power-law index of 1/4 is smaller than that of Equation (\ref{eq:eq_nonlinearity_beta}), it is generally expected that larger maximum nonlinearity of Alfv\'en wave is possible in higher-$\beta$ stellar wind.

Because $\beta_{\phi\max}\propto c_{s,\phi\max}^2/V_{A\phi\max}^2\propto T_{\phi\max}/V_{A\phi\max}^2$ ($c_{s,\phi\max}$ is the sound speed at the distance where Alfv\'en wave amplitude reaches a maximum) and $T_{\phi\max}\sim10^6$ K regardless of the stars, it is expected that $V_{A\phi\max}\propto\beta^{-1/2}$. Our simulation results show
\begin{equation}
  V_{A\phi\max}=2.15\times10^3\ \mbox{km s$^{-1}$}\left(\beta_{\phi\max}\over 10^{-2}\right)^{-0.549},
\end{equation}
and consequently, we obtain
\begin{equation}
  v_{\phi\max}=286\ \mbox{km s$^{-1}$}\left(\beta_{\phi\max}\over10^{-2}\right)^{-0.171}.
  \label{eq:eq_vphimax_beta}
\end{equation}

Equations (\ref{eq:vwind_estimate}) and (\ref{eq:eq_vphimax_beta}) determine $v_{\rm wind}$ as a function of $\beta_{\phi\max}$. In particular, lower $\beta_{\phi\max}$ leads to faster $v_{\rm wind}$, because larger-amplitude Alfv\'en wave can propagate in the stellar wind.

\subsection{Coronal Temperature and  Mass-loss rate}
\label{sec:tco_empirical}

We expect that $T_{\rm co}$ (coronal temperature at $r=1.1r_\star$) is expressed as a function of $F_{A,\rm co}$ (transmitted Alfv\'en wave energy flux into the corona at $r=1.1r_\star$) through the energy conservation law (Equation (\ref{eq:energy_conservation}) and Equation (\ref{eq:fcco_vs_tco}). Here, the energy conservation law is simplified as below.
\begin{equation}
  -L_{c,\rm co}\approx L_{A,\rm co}-(L_{\rm kin,wind}-L_{g,\rm co}).
  \label{eq:energy_conservation_app_2}
\end{equation}
Note that the radiative energy loss, which is neglected in the above, can be involved with energy conservation when the open flux tube filling factor ($f_{\rm ph}$) is much smaller than that used in this study (1/1600). The detail is described in Appendix \ref{sec:appendix_fph}.

Because
\begin{equation}
  L_{\rm kin,wind}=\dot{M}{v_{\rm wind}^2\over2},
\end{equation}
and
\begin{equation}
  L_{g,\rm co}=-\dot{M}{v_{\rm esc\star}^2/2}=-\dot{M}{GM_\star/r_\star},
\end{equation}
we obtain the following by dividing the both sides of Equation (\ref{eq:energy_conservation_app_2}) with $L_{A,\rm co}$.
\begin{equation}
  \alpha_{c/A}=1-\alpha_{{\rm wind}/A}\left(1+{v_{\rm esc\star}^2\over v_{\rm wind}^2}\right),
  \label{eq:alpha_c_a}
\end{equation}
where $\alpha_{c/A}$ and $\alpha_{{\rm wind}/A}$ are the energy conversion efficiencies from the Alfv\'en wave energy flux ($L_{A,\rm co}$) to the heat conduction flux ($L_{c,\rm co}$) and wind's kinetic energy flux ($L_{\rm kin,wind}$), respectively. That means
\begin{equation}
  \alpha_{c/A}=-L_{c,\rm co}/L_{A,\rm co},
\end{equation}
and
\begin{equation}
  \alpha_{{\rm wind}/A}=L_{\rm kin,wind}/L_{A,\rm co}.
\end{equation}

\begin{figure}
  \begin{center}
    \epsscale{1.}
    \plotone{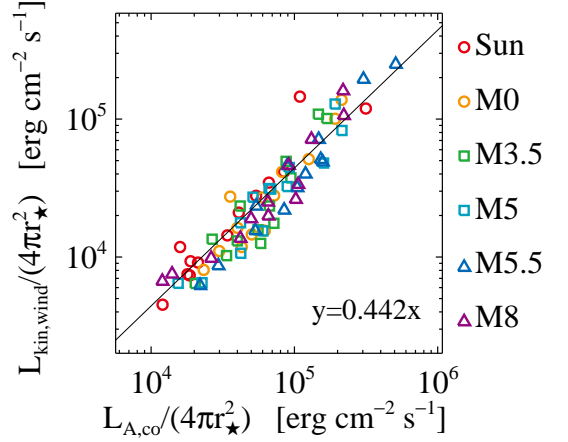}
    \caption{The transmitted Alfv\'en wave energy flux into the corona ($L_{A,\rm co}$) vs stellar wind's kinetic energy flux ($L_{\rm kin,wind}$).}
    \label{fig:figure14.eps}
  \end{center}
\end{figure}

Hereafter, we assume that $\alpha_{{\rm wind}/A}$ is independent of stars, chromospheric magnetic field strengths, and energy inputs from the photosphere; namely $\alpha_{{\rm wind}/A}=0.442\pm0.166$. This assumption is confirmed at least in our parameter survey (Figure \ref{fig:figure14.eps}), and enables us to solve the degeneracy between the coronal temperature ($T_{\rm co}$) and mass-loss rate ($\dot{M}$) in the energy conservation law. That means, by using $\alpha_{{\rm wind}/A}$, we can express $T_{\rm co}$ and $\dot{M}$ as the functions of $F_{A,\rm co}$ as follows.
\begin{equation}
  T_{\rm co}=a_2\left[\left\{1-\alpha_{{\rm wind}/A}\left(1+{v_{\rm esc\star}^2\over v_{\rm wind}^2}\right)\right\}\tilde{F}_{A,\rm co}\tilde{l}_{B,\rm co}\right]^{k_2},
  \label{eq:eq_t_cr_final}
\end{equation}
where $\tilde{F}_{A,\rm co}=F_{A,\rm co}/(10^5$ erg cm$^{-2}$ s$^{-1})$, $\tilde{l}_{B,\rm co}=l_{B,\rm co}/r_\odot$. $a_2$ and $k_2$ are determined by our simulation results; $a_2=1.62\times10^6$ K, $k_2=0.256$ (see Equation (\ref{eq:fcco_vs_tco})).
\begin{equation}
  \dot{M}=2\alpha_{{\rm wind}/A}{L_{A,\rm co}\over v_{\rm wind}^2}.
  \label{eq:mdot_la_v^2}
\end{equation}

It should be noted that, however, $\alpha_{{\rm wind}/A}$ possibly depends on the filling factor of open flux tube ($f_{\rm ph}$), which is beyond our present parameter survey.
In addition, $\alpha_{c/A}$ of Equation (\ref{eq:alpha_c_a}) is often quenched to zero when $v_{\rm wind}<v_{\rm esc\star}$, which means the approximation for Equations (\ref{eq:energy_conservation_app}) and (\ref{eq:alpha_c_a}) become invalid. In the following analysis, therefore, we assume the monotonic increase in $|L_{c,\rm co}|$ with $L_{A,\rm co}$. i.e., $\partial\ln\alpha_{c/A}/\partial\ln L_{A,\rm co}>-1$ to avoid this problem.

\subsection{Plasma $\beta$ of Stellar Wind}
\label{sec:beta_empirical}
\begin{figure}
  \begin{center}
    \epsscale{1}
    \plotone{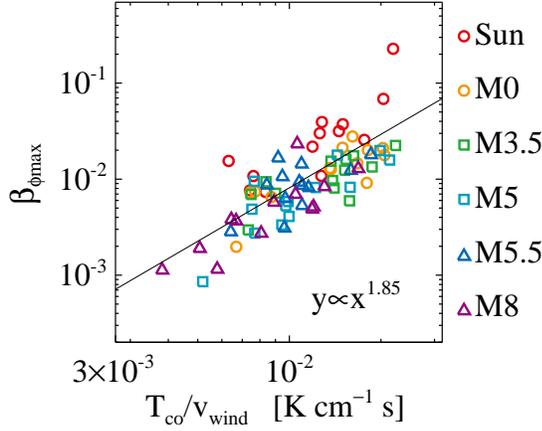}
    \caption{The plasma $\beta$ at the distance where the amplitude of Alfv\'en wave reaches maximum ($\beta_{\phi\max}$) as a function of $T_{\rm co}/v_{\rm wind}$. The solid line corresponds to Equation (\ref{eq:beta_scale}).}
    \label{fig:figure15.eps}
  \end{center}
\end{figure}

In Section \ref{sec:vwind_empirical}, we express $v_{\rm wind}$ as a function of $\beta_{\phi\max}$. $\beta_{\phi\max}$ is the plasma $\beta$ at the distance (hereafter, $r_{\phi\max}$) where the amplitude of Alfv\'en wave reaches the maximum. In this section, thus, we consider how $\beta_{\phi\max}$ is determined. The physical quantities with the subscript $_{\phi\max}$ means those at $r=r_{\phi\max}$ in the following.

The plasma $\beta$ of stellar wind is expressed as below.
\begin{equation}
  \beta
  ={8\pi p\over B_x^2}
  ={8\pi(\gamma-1)\over\gamma v_x}{L_{\rm ent}/(4\pi fr_\star^2)\over (B_{\rm ph}f_{\rm ph}/f)^2}\left(r\over r_\star\right)^2,
  \label{eq:beta_estimate}
\end{equation}
where $L_{\rm ent}=4\pi fr^2\gamma pv_x/(\gamma-1)$ is the enthalpy luminosity. In the isothermal stellar wind, $L_{\rm ent}=\gamma R_g\dot{M}T/\{\mu(\gamma-1)\}$ is almost constant, and then $\beta_{\phi\max}\propto (r_{\phi\max}/r_\star)^2/v_{x,\phi\max}$. In the following, we aim at expressing $\beta_{\phi\max}$ with several integral constants, such as $\dot{M}$, $B_{\rm ph}f_{\rm ph}$, $v_{\rm wind}$, and coronal parameters, such as $T_{\rm co}$, $L_{A,\rm co}$.

First, $r/r_\star$ in Equation (\ref{eq:beta_estimate}) can be related to the ratios of mass density, Alfv\'en wave energy luminosity ($L_A\approx\sqrt{4\pi\rho}B_x fr^2v_\phi^2$), and Alfv\'en wave nonlinearity ($\eta=v_\phi/V_{Ax}$). In fact, by using $B_xfr^2=$const., we have $L_A\propto\sqrt{\rho}v_\phi^2$ and $v_{\phi}\propto\eta/(\sqrt{\rho}fr^2)$, and then,
\begin{equation}
  {L_A\over L_{A,\rm co}}
  =\sqrt{\rho_{\rm co}\over\rho}
  \left(\eta\over\eta_{\rm co}\right)^2
  \left(f_{\rm co}r_{\rm co}^2\over fr^2\right)^2,
  \label{eq:L_A_conservation}
\end{equation}
where $f_{\rm co}$ is the filling factor of open flux tube at $r=r_{\rm co}=1.1r_\star$.

$\sqrt{\rho_{\rm co}/\rho}$ in Equation (\ref{eq:L_A_conservation}) is inconvenient for later discussion, and rewritten as follows
\begin{equation}
  \sqrt{\rho_{\rm co}\over\rho}={M_{Ax}\over M_{Ax,\rm co}}={v_x\over v_\phi}{\eta\over M_{Ax,\rm co}}.
\end{equation}
Note that $M_{Ax}$ is Alfv\'en Mach number of stellar wind, which is expressed as $M_{Ax}=v_x/V_{Ax}=\dot{M}/(\sqrt{4\pi\rho}Bfr^2)\propto\rho^{-1/2}$.

By substituting the above into Equation (\ref{eq:L_A_conservation}), we have
\begin{equation}
  \left(r\over r_{\rm co}\right)^4
  ={L_{A,\rm co}\over L_A}{\eta^3\over\eta_{\rm co}^2M_{Ax,\rm co}}{v_x\over v_\phi}
  \left(f_{\rm co}\over f\right)^2.
  \label{eq:rphimax_rco^4}
\end{equation}

$\eta^2_{\rm co}M_{Ax,\rm co}$ in the above can be further rewritten as follows.
\begin{align}
  \eta_{\rm co}^2={v_{\phi,\rm co}^2\over V_{Ax,\rm co}^2}
  =&{1\over V_{Ax,\rm co}^2}{L_{A,\rm co}\over\sqrt{4\pi\rho_{\rm co}}B_{x,\rm co}f_{\rm co}r_{\rm co}^2}\nonumber\\
  =&{4\pi\rho_{\rm co}\over B_{x,\rm co}^2}{L_{A,\rm co}\over\sqrt{4\pi\rho_{\rm co}}B_{\rm ph}f_{\rm ph}r_\star^2}\nonumber\\
  =&\sqrt{4\pi\rho_{\rm co}}{L_{A,\rm co}(f_{\rm co}r_{\rm co}^2)^2\over(B_{\rm ph}f_{\rm ph}r_\star^2)^3}
\end{align}
\begin{align}
  M_{Ax,\rm co}=&{v_{x,\rm co}\over V_{Ax,\rm co}}={v_{x,\rm co}f_{\rm co}r_{\rm co}^2\over V_{Ax,\rm co}f_{\rm co}r_{\rm co}^2}\nonumber\\
  =&{\dot{M}/(4\pi\rho_{\rm co})\over B_{\rm co}f_{\rm co}r_{\rm co}^2/\sqrt{4\pi\rho_{\rm co}}}\nonumber\\
  =&{\dot{M}\over\sqrt{4\pi\rho_{\rm co}}}{1\over B_{\rm ph}f_{\rm ph}r_{\rm ph}^2}
\end{align}


These lead to
\begin{equation}
  \eta_{\rm co}^2M_{Ax,\rm co}={1\over(B_{\rm ph}f_{\rm ph})^4}\left(r_{\rm co}\over r_\star\right)^4{L_{A,\rm co}\over r_\star^2}{\dot{M}\over r_\star^2}f_{\rm co}^2.
  \label{eq:nonlinearity_corona}
\end{equation}

By substituting the above into Equation (\ref{eq:rphimax_rco^4}),
\begin{equation}
  \left(r\over r_\star\right)^4
  =\eta^3{(B_{\rm ph}f_{\rm ph})^4r_\star^4\over\dot{M}L_A}{v_x\over v_\phi}f^{-2},
  \label{eq:rphimax_rstar}
\end{equation}

Combining the above and Equation (\ref{eq:beta_estimate}) leads to
\begin{equation}
  \beta
  ={2R_g\over\mu}\sqrt{\dot{M}\over L_A}T{\eta^{3/2}\over v_\phi}\sqrt{v_\phi\over v_x}
  \label{eq:beta_etimate1}
\end{equation}

$\beta_{\phi\max}$ is obtained by using $L_A=L_{A,\phi\max}$, $T=T_{\phi\max}\approx T_{\rm co}$, $\eta=\eta_{\phi\max}$, $v_\phi=v_{\phi\max}$, and $v_x=v_{x,\phi\max}$.

Meanwhile, we can expect the following three. First, $\sqrt{\dot{M}/L_{A,\phi\max}}\propto 1/v_{\rm wind}$ because $L_{A,\phi\max}\propto L_{A,\rm co}$ and Equation (\ref{eq:mdot_la_v^2}).
Our parameter survey shows $L_{A,\phi\max}/L_{A,\rm co}=0.193\pm0.087$. Second, $\eta_{\phi\max}$ and $v_{\phi\max}$ are written as the power-law functions of $\beta_{\phi\max}$ (Equations (\ref{eq:eq_nonlinearity_beta}) and (\ref{eq:eq_vphimax_beta})). Third, $v_{x,\phi\max}\propto v_{\phi\max}$ because $v_{x,\phi\max}$ is determined by $v_{\phi\max}$ in the  Alfv\'en wave driven wind. Our parameter survey shows $v_{x,\phi\max}/v_{\phi\max}=1.35\pm0.27$.

Based on these considerations, we assume the following power-law relation between $\beta_{\phi\max}$ and $T_{\rm co}/v_{\rm wind}$.
\begin{equation}
  \beta_{\phi\max}=a_3\left(\tilde{T}_{\rm co}\over\tilde{v}_{\rm wind}\right)^{k_3},
  \label{eq:beta_scale}
\end{equation}
where $\tilde{T}_{\rm co}=T_{\rm co}/10^6$ K, $\tilde{v}_{\rm wind}=v_{\rm wind}/600$ km s$^{-1}$, and $a_3$ and $k_3$ are determined by our simulation results. $a_3=2.09\times10^{-2}$ and $k_3=1.85$ (Figure \ref{fig:figure15.eps}).

It is notable that $\beta_{\phi\max}$ is not explicitly dependent on the stellar magnetic field strength ($B_{\rm ph}f_{\rm ph}$). Although the plasma $\beta$ at a given distance is negatively correlated with $B_{\rm ph}f_{\rm ph}$ (Equation (\ref{eq:beta_estimate})), the distance where the amplitude of Alfv\'en wave reaches the maximum ($r_{\phi\max}$) is positively correlated with $B_{\rm ph}f_{\rm ph}$ (Equation (\ref{eq:rphimax_rstar})) because the nonlinearity of Alfv\'en wave at a given distance tends to be smaller when $B_{\rm ph}f_{\rm ph}$ becomes larger (Equation (\ref{eq:nonlinearity_corona})). These effects of varying $B_{\rm ph}f_{\rm ph}$ are canceled out with each other in Equation (\ref{eq:beta_scale}).


\subsection{Transmissivity of Alfv\'en Wave into the Stellar Corona}
\label{sec:transmissivity}

\begin{figure}
  \begin{center}
    \epsscale{1}
    \plotone{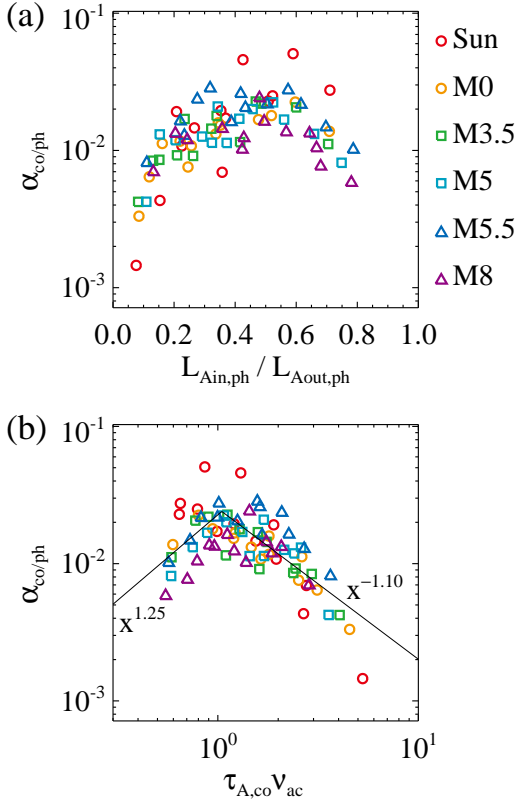}
    \caption{(a) The relation between transmissivity of Alfv\'en wave ($\alpha_{\rm co/ph}=L_{A,\rm co}/L_{A\rm out,ph}$) and reflection rate of the photosphere ($L_{A\rm in,ph}/L_{A\rm out,ph}$). (b) The relation between $\alpha_{\rm co/ph}$ and the normalized Alfv\'en wave travel time from the photosphere to the corona ($\tau_{A,\rm co}$ defined in Equation (\ref{eq:def_tau_A})) with respect to the acoustic cutoff frequency $\nu_{\rm ac}$.}
    \label{fig:figure16.eps}
  \end{center}
\end{figure}

From Section \ref{sec:vwind_empirical} to \ref{sec:beta_empirical}, we describe the relations among $v_{\rm wind}$, $T_{\rm co}$, and $\beta_{\phi\max}$. In particular, Equations (\ref{eq:vwind_estimate}), (\ref{eq:eq_t_cr_final}) and (\ref{eq:beta_scale}) can be used to estimate $v_{\rm wind}$ and $T_{\rm co}$ for a given parameter set of $v_{\rm esc\star}$, $l_{B,\rm co}$ and $F_{A,\rm co}$.

$F_{A,\rm co}$ is the transmitted Alfv\'en wave energy flux into the corona, and related to Alfv\'en wave energy input from the photosphere ($L_{A\rm out,ph}$). Here,
\begin{equation}
  L_{A\rm out/in}=4\pi r^2fF_{A\rm out/in}
\end{equation}
corresponds to the luminosity of outward/inward Alfv\'en wave. $F_{A,\rm out/in}$ is the Poynting flux associated with the outward/inward Alfv\'en wave:
\begin{equation}
  F_{A\rm out/in}={1\over4}\rho z_{\rm out/in}^2,
\end{equation}
where
\begin{equation}
  z_{\rm out}=v_\phi-{B_\phi\over\sqrt{4\pi\rho}},
\end{equation}
and
\begin{equation}
  z_{\rm in}=v_\phi+{B_\phi\over\sqrt{4\pi\rho}}.
\end{equation}

We discuss the transmissivity of Alfv\'en wave from the stellar photosphere to corona by defining it as $\alpha_{\rm co/ph}=L_{A,\rm co}/L_{A\rm out, ph}$ in the following.

Figure \ref{fig:figure16.eps}(a) shows the relation between $\alpha_{\rm co/ph}$ and $L_{A\rm in, ph}/L_{A\rm out, ph}$. The latter represents the fraction of Alfv\'en wave energy reflected back to the stellar photosphere. As seen in this panel, while larger reflection rate than $\sim0.5$ is negatively correlated with transmissivity of Alfv\'en wave, the correlation is reversed when the reflection rate is smaller than $\sim0.5$.
This positive correlation between the transmissivity and reflection rate of Alfv\'en wave suggests that the Alfv\'en wave energy is attenuated mainly due to the wave dissipation, rather than reflection. To characterize this wave dissipation, we introduce the Alfv\'en travel time from the photosphere to corona ($\tau_{A,\rm co}$) as below.
\begin{equation}
  \tau_{A,\rm co}=\int^{1.1r_\star}_{r_\star}{1\over B_x/\sqrt{4\pi\rho}}{dx\over dr}dr
  \label{eq:def_tau_A}
\end{equation}

Figure \ref{fig:figure16.eps}(b) shows the correlation between $\alpha_{\rm co/ph}$ and the nondimensionalized Alfv\'en travel time with respect to the typical wave frequency $\nu_A$, where we take the acoustic-cutoff frequency of the photosphere ($\nu_{\rm ac}$ in Table \ref{table:table_stars}) for $\nu_A$. Figure \ref{fig:figure16.eps}(b) looks like Figure \ref{fig:figure16.eps}(a) but they are flipped left and right.
When $\tau_{A,\rm co}\nu_{\rm ac}\gtrsim1$, the chromosphere is too thick for Alfv\'en wave to transmit into the corona. The resultant transmissivity of Alfv\'en wave practically follows
\begin{equation}
  \alpha_{\rm co/ph}\approx a_{4,1}\left(\tau_{A,\rm co}\nu_{\rm ac}\over a_{4,2}\right)^{k_4},
\end{equation}
where $a_{4,1}=2.41\times10^{-2}$, $a_{4,2}=1.04$, and 
\begin{equation}
  k_4=
  \left\{
  \begin{array}{ll}
    1.25 & (\tau_{A,\rm co}\nu_{\rm ac}<a_{4,2})\\
    -1.10 & (\tau_{A,\rm co}\nu_{\rm ac}>a_{4,2})
  \end{array}
  \right.
\end{equation}

It should be noted that $\tau_{A,\rm co}$ is approximated by the Alfv\'en travel time up to the merging height ($H_m$, the height at which $B_x=\overline{B}$). This is because the integrand in Equation (\ref{eq:def_tau_A}) decreases exponentially with increasing Alfv\'en speed above $H_m$. That means we can expect that $\tau_{A,\rm co}\nu_{\rm ac}\approx (x_m/\overline{V}_A)\nu_{\rm ac}$, where
\begin{equation}
  x_m=\int^{r_\star+H_m}_{r_\star}{dx\over dr}dr,
\end{equation}
\begin{equation}
  \overline{V}_A={1\over x_m}\int^{r_\star+H_m}_{r_\star}{B_x\over\sqrt{4\pi\rho}}{dx\over dr}dr.
\end{equation}
Our parameter survey shows $\tau_{A,\rm co}\nu_{\rm ac}\approx 1.96(x_m\nu_{\rm ac}/\overline{V}_A)^{1.17}$. Furthermore, $x_m\nu_{\rm ac}=$a factor $\times c_{s,\rm ph}$, depending on $\overline{B}$, and $\overline{V}_A\propto c_{s,\rm ph}$ in this study. $\overline{V}_A$ is also influenced on by the Alfv\'en wave amplitude of the photosphere ($v_{\rm ph}$) through the varying density profile below $H_m$. These considerations lead to the following practical fit.
\begin{equation}
  \tau_{A,\rm co}\nu_{\rm ac}\approx a_5\tilde{g}_\star^{k_{5,1}}\left(\overline{B}\over B_{\rm ph}\right)^{-k_{5,2}}\left(v_{\rm ph}\over c_{s,\rm ph}\right)^{k_{5,3}},
  \label{eq:tau_A_nu_ac_scale}
\end{equation}
where $\tilde{g}_\star=g_\star/10^5$ cm s$^{-2}$, $a_5=0.921$, $k_{5,1}=0.240$, $k_{5,2}=0.408$ and $k_{5,3}=0.697$.

Equation (\ref{eq:tau_A_nu_ac_scale}) shows that the weaker magnetic field is, the larger $\tau_{A,\rm co}\nu_{\rm ac}$ and the more significant wave dissipation is in the lower atmosphere. 
When $\tau_{A,\rm co}\nu_{\rm ac}\lesssim1$, on the other hand, Alfv\'en wave transmits into the corona in shorter time scale than its typical wave period. In this case, the wave dissipation is relatively negligible compared to the wave reflection, resulting in the negative correlation between the reflection rate and transmissivity of Alfv\'en wave.


\subsection{Plasma Pressure at Transition Region}
\label{sec:ptr_empirical}

The plasma pressure at transition region ($p_{\rm tr}$) is determined so that the radiative cooling approximately balances with the heat conduction heating.
\begin{equation}
  n_{\rm tr}^2\Lambda(T_{\rm tr})\approx|{\rm div}F_c|_{\rm tr},
\end{equation}
where $n=\rho/m_p$, and $\Lambda(T)$ represents radiative loss function for the optically thin plasma. By substituting $n_{\rm tr}=p_{\rm tr}/(2k_BT_{\rm tr})$, where $T_{\rm tr}=4\times10^4$ K and $k_B$ is the Boltzmann constant, we obtain
\begin{equation}
  p_{\rm tr}^2\approx{(2k_BT_{\rm tr})^2\over\Lambda(T_{\rm tr})}|{\rm div}F_c|_{\rm tr}.
\end{equation}

\begin{figure}
  \begin{center}
    \epsscale{1.}
    \plotone{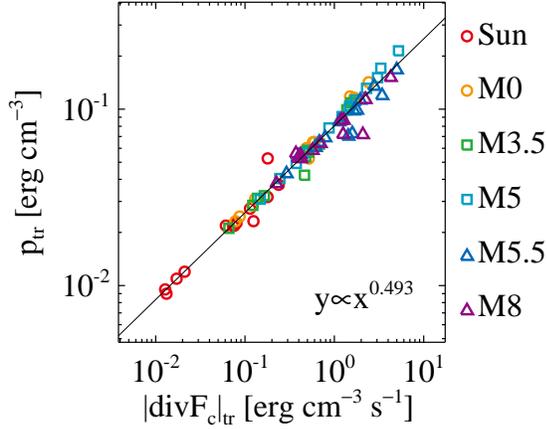}
    \caption{The relation between plasma pressure at transition region $p_{\rm tr}$ and heating rate due to the heat conduction flux $|{\rm div}F_c|_{\rm tr}$. The strong positive correlation results from the energy balance between radiative cooling and heat conduction heating at transition region.}
    \label{fig:figure17.eps}
  \end{center}
\end{figure}

The above estimate is consistent with the simulation results shown in Figure \ref{fig:figure17.eps}, namely:
\begin{equation}
  p_{\rm tr}=8.05\times10^{-2}\ \mbox{erg cm$^{-3}$}\times\left(|{\rm div}F_c|_{\rm tr}\over10^{-2}\ \mbox{erg cm$^{-3}$ s$^{-1}$}\right)^{0.493}.
  \label{eq:eq_p_tr_0}
\end{equation}

We expect that $|{\rm div}F_c|_{\rm tr}$ is characterized with the coronal temperature ($T_{\rm co}$) and the spatial scale of expanding magnetic flux tube at transition region ($l_{B,\rm tr}$). Because $L_{c,\rm tr}\propto L_{c,\rm co}$ is rewritten by $A_{\rm tr}T_{\rm tr}^{7/2}/l_{B,\rm tr}\propto A_{\rm co}T_{\rm co}^{7/2}/l_{B,\rm co}$, it is appropriate that $l_{B,\rm tr}$ is defined with $l_{B,\rm co}$ in Equation (\ref{eq:l_b_definition}) as below.
\begin{equation}
  l_{B,\rm tr}={A_{\rm tr}\over A_{\rm co}}l_{B,\rm co}=\int^{x_{\rm co}}_{x_{\rm tr}}dx{A_{\rm tr}\over A}.
  \label{eq:l_b_tr_definition}
\end{equation}

In fact, the regression analysis clearly shows 
\begin{equation}
  |{\rm div}F_c|_{\rm tr}=1.35\times10^{-5}\ \mbox{erg cm$^{-3}$ s$^{-1}$}\times\tilde{l}_{B,\rm tr}^{-1.70}\tilde{T}_{\rm co}^{4.07},
  \label{eq:eq_divfc_0}
\end{equation}
where $\tilde{l}_{B,\rm tr}=l_{B,\rm tr}/r_\odot$ and $\tilde{T}_{\rm co}=T_{\rm co}/10^6$ K.

From Equation (\ref{eq:eq_p_tr_0}) and (\ref{eq:eq_divfc_0}), we obtain
\begin{equation}
  p_{\rm tr}=3.31\times10^{-3}\ \mbox{erg cm$^{-3}$}\times\tilde{l}_{B,\rm tr}^{-0.824}\tilde{T}_{\rm co}^{2.07}.
  \label{eq:eq_p_tr_T_cr}
\end{equation}

For the comparison, we note the prediction by RTV scaling (Section \ref{sec:p_tr})
\begin{equation}
  p_{\rm loop}=5.24\times10^{-2}\ \mbox{erg cm$^{-3}$}\times\left(l\over r_\odot\right)^{-1}\left(T_{\rm loop}\over10^6\ \mbox{K}\right)^3,
\end{equation}
where the plasma pressure in the coronal loop ($p_{\rm loop}$) is related to the coronal loop temperature ($T_{\rm loop}$) and loop length ($l$).

By substituting Equation (\ref{eq:eq_t_cr_final}) into Equation (\ref{eq:eq_p_tr_T_cr}) and using Equation (\ref{eq:l_b_tr_definition}), we obtain
\begin{align}
  p_{\rm tr}=&9.00\times10^{-3}\ \mbox{erg cm$^{-3}$}\nonumber\\
  &\times\tilde{l}_{B,\rm tr}^{-0.294}\left(A_{\rm co}\over A_{\rm tr}\right)^{0.529}(\alpha_{c/A}\tilde{F}_{A,\rm co})^{0.529},
  \label{eq:eq_p_tr_final}
\end{align}
where $\alpha_{c/A}$ is defined by Equation (\ref{eq:alpha_c_a}) and $\tilde{F}_{A,\rm co}=F_{A,\rm co}/10^5$ erg cm$^{-2}$ s$^{-1}$.

The coronal mass density $\rho_{\rm co}$ is immediately obtained from Equation (\ref{eq:eq_p_tr_final}). By assuming the hydrostatic atmosphere with the constant gravitational acceleration of $g_\star$, the pressure at $r=1.1r_\star$ ($p_{\rm co}$) is estimated as $p_{\rm co}\approx e^{-T_\star/T_{\rm co}}p_{\rm tr}$, where 
\begin{equation}
  T_\star=0.1r_\star\mu g_\star/R_g\approx1.15\times10^6 \mbox{ K } \left(M_\star/r_\star\over M_\odot/r_\odot\right).
\end{equation}
The ratio of $T_\star/T_{\rm co}$ is identical to $0.1r_\star/H_{p,\rm co}$, where $H_{p,\rm co}=R_gT_{\rm co}/(\mu g_\star)$ is the pressure scale height in the corona.
By using Equation (\ref{eq:eq_p_tr_final}) and $\rho_{\rm co}=\mu p_{\rm co}/R_gT_{\rm co}$, we obtain
\begin{align}
  \rho_{\rm co}=&3.33\times10^{-17}\mbox{g cm$^{-3}$}\nonumber\\
  &\times e^{-T_\star/T_{\rm co}}\tilde{l}_{B,\rm tr}^{-0.550}\left(A_{\rm co}\over A_{\rm tr}\right)^{0.273}(\alpha_{c/A}\tilde{F}_{A,\rm co})^{0.273}
  \label{eq:eq_rho_cr_final}
\end{align}

  As suggested in Equation (\ref{eq:eq_t_cr_final}), the coronal temperature in M dwarfs with smaller $l_{B,\rm co}$ is cooler than that of the Sun for a given $F_{A,\rm co}$. This is because the temperature profile in the corona with smaller $l_{B,\rm co}$ is characterized with larger temperature gradient. In this case, the hotter coronal temperature is not required for the heat conduction flux from the corona to balance with the transmitted Alfv\'en wave energy flux into the corona.

  $p_{\rm tr}$ is less dependent of $l_{B,\rm tr}$ or $l_{B,\rm co}$ for a given $F_{A,\rm co}$. This is because the cooler coronal temperature as a result of smaller $l_{B,\rm co}$ cancels out the tendency that smaller $l_{B,\rm tr}$ leads to larger temperature gradient around the transition region. The simulated $p_{\rm tr}$ of M dwarf is, on the other hand, relatively higher than solar value (e.g., Figure \ref{fig:figure17.eps}). This is partly because $F_{A,\rm co}$ of M dwarf tends to be larger than that of the Sun due to the higher transmissivity of Alfv\'en wave across the stellar chromosphere. It should be noted that the plasma pressure of the photosphere ($p_{\rm ph}$) more largely increases with decreasing $T_{\rm eff}$ compared to the increase in $p_{\rm tr}$. As a result, $p_{\rm tr}/p_{\rm ph}$ decreases with decreasing $T_{\rm eff}$, leading to more extended chromosphere associated with higher $H_{\rm tr}/H_{\rm ph}$ (Figure \ref{fig:figure06.eps}).

\begin{figure*}
  \begin{center}
    \epsscale{1.2}
    \plotone{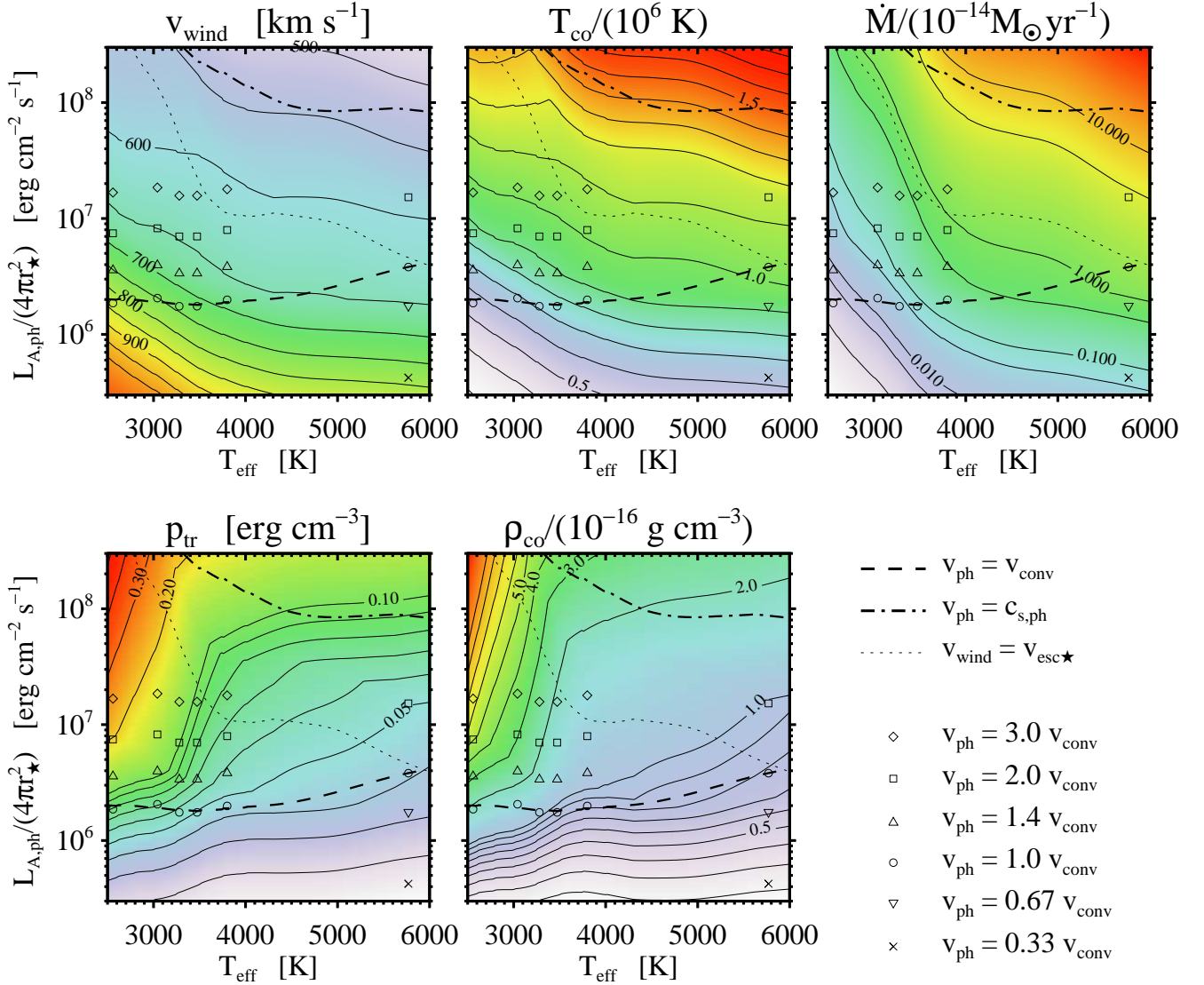}
    \caption{$T_{\rm eff}$-$L_{A,\rm ph}$ diagrams for $T_{\rm co}$, $v_{\rm wind}$, $\dot{M}$, $p_{\rm tr}$ and $\rho_{\rm co}$.
      The symbols, such as diamonds, squares, triangles, represent the positions of $(T_{\rm eff},L_{A,\rm ph})$ in our parameter survey about the Alfv\'en wave amplitude on the photosphere ($v_{\rm ph}$). Note that these diagrams are drawn by specifying $\overline{B}/B_{\rm ph}=e^{-4}$.
      The thick dashed line corresponds to the fiducial $L_{A,\rm ph}$ as a function of $T_{\rm eff}$, which is the case that $v_{\rm ph}$ is equal to the convective velocity ($v_{\rm conv}$). The thick dashed–dotted line corresponds to the largest $L_{A,\rm ph}$ obtained by assuming that the convective velocity reaches the sound speed of the photosphere (i.e., $v_{\rm ph}=c_{s,\rm ph}$). The thin dashed line represents $L_{A,\rm ph}$ as a function of $T_{\rm eff}$, which results in $v_{\rm wind}=v_{\rm esc\star}$.}
    \label{fig:figure18.eps}
  \end{center}
\end{figure*}

\subsection{$T_{\rm eff}$-$L_{A,\rm ph}$ Diagrams for M-Dwarfs' Atmospheres and Winds}
\label{sec:h-r_diagram}

On the basis of discussion in Section \ref{sec:transmissivity}, we can calculate the transmitted Alfv\'en wave energy flux into the corona ($F_{A,\rm co}$) by specifying the Alfv\'en wave energy input from the photosphere ($L_{A,\rm ph}$), basic parameters of stars such as $r_\star$, $M_\star$, $T_{\rm eff}$, and the parameters of open flux tube configuration such as $B_{\rm ph}$, $\overline{B}$, $l_{B,\rm co}$. The obtained $F_{A,\rm co}$ is necessary to predict the stellar coronal temperature ($T_{\rm co}$), stellar wind velocity ($v_{\rm wind}$) and mass-loss rate ($\dot{M}$), according to Section \ref{sec:vwind_empirical} to \ref{sec:beta_empirical}. Furthermore, the plasma pressure at transition region ($p_{\rm tr}$) and coronal mass density ($\rho_{\rm co}$) can be derived by using $T_{\rm co}$ (Section \ref{sec:ptr_empirical}).

Therefore, when we limit our interest to the main-sequence stars' atmospheres and winds so that the parameters of $r_\star$, $M_\star$, $B_{\rm ph}$, $\overline{B}$, $l_{B,\rm co}$ can be roughly expressed as the functions of $T_{\rm eff}$, it is possible to estimate these physical quantities for given $T_{\rm eff}$ and $L_{A,\rm ph}$.
We developed a python code called AWSAWS to calculate $v_{\rm wind}$, $T_{\rm co}$, $\dot{M}$, $p_{\rm tr}$, $\rho_{\rm co}$ as functions of $T_{\rm eff}$ and $L_{A,\rm ph}$. The code is provided on the first author's Web site\footnote{\url{https://www.kwasan.kyoto-u.ac.jp/\%7Esakaue/awsaws/awsaws.py}}.

The resultant $T_{\rm eff}$-$L_{A,\rm ph}$ diagrams for $v_{\rm wind}$, $T_{\rm co}$, $\dot{M}$, $p_{\rm tr}$ and $\rho_{\rm co}$ are drawn in Figure \ref{fig:figure18.eps}. Note that we assume $T_{\rm eff}$-$r_\star$ and $r_\star$-$M_\star$ relations of main-sequence stars as explained in Appendix \ref{sec:MS_r_M}. $\overline{B}/B_{\rm ph}=e^{-4}$ is also assumed to obtain Figure \ref{fig:figure18.eps}.
  The chromospheric magnetic field strength ($\overline{B}$) can have an influence on the physical quantities in Figure \ref{fig:figure18.eps} by a factor of a few, through the transmissivity of Alfv\'en wave energy from the photosphere to the corona (Section \ref{sec:transmissivity}, especially Equation (\ref{eq:tau_A_nu_ac_scale})). The overall picture of Figure \ref{fig:figure18.eps} is, however, still identical enough to discuss the trends of physical quantities, even by changing $\overline{B}$.

In Figure \ref{fig:figure18.eps}, the thick dashed line corresponds to the fiducial $L_{A,\rm ph}$ as a function of $T_{\rm eff}$, which is the case that Alfv\'en wave amplitude of the photosphere ($v_{\rm ph}$) is equal to the convective velocity ($v_{\rm conv}$). The thick dashed–dotted line corresponds to the largest $L_{A,\rm ph}$ obtained by assuming that the convective velocity reaches the sound speed of the photosphere (i.e., $v_{\rm ph}=c_{s,\rm ph}$). The thin dashed line represents $L_{A,\rm ph}$ as a function of $T_{\rm eff}$, which results in $v_{\rm wind}=v_{\rm esc\star}$.

Along the thick dashed line, it is seen that $v_{\rm wind}$ and $T_{\rm co}$ are faster and cooler with decreasing $T_{\rm eff}$, and that the mass-loss rate $\dot{M}$ of M dwarfs' winds are much smaller than the solar wind's value. The differences in $p_{\rm tr}$ and $\rho_{\rm co}$ are less remarkable, but they systematically increase with decreasing $T_{\rm eff}$, as discussed in Section \ref{sec:ptr_empirical}.

\begin{figure*}
  \begin{center}
    \epsscale{1.}
    \plotone{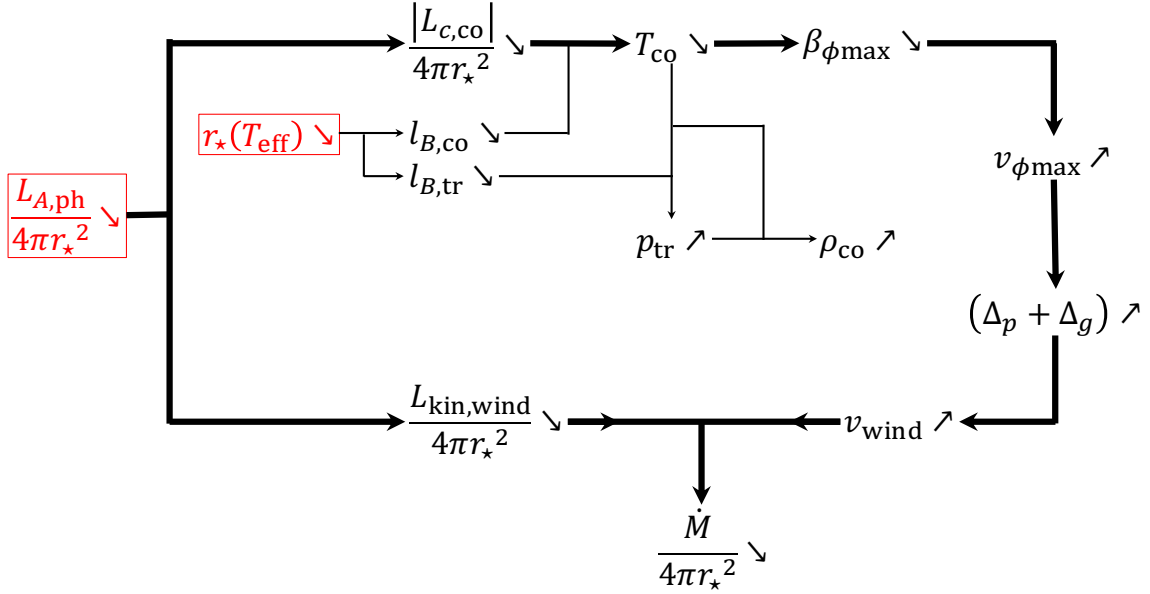}
    \caption{The causal relations among the varying $v_{\rm wind}$, $T_{\rm co}$, $\dot{M}$, $p_{\rm tr}$ and $\rho_{\rm co}$ with respect to decreases in $L_{A,\rm ph}/(4\pi r_\star^2)$ and $r_\star(T_{\rm eff})$. Here, $L_{A,\rm ph}/(4\pi r_\star^2)$ and $r_\star(T_{\rm eff})$ correspond to the vertical and horizontal axes of panels in Figure \ref{fig:figure18.eps}, respectively.
      The thick arrows depict the results of energy partitioning from $L_{A,\rm ph}$ to $L_{c,\rm co}$ or $L_{\rm kin,wind}$.
      In particular, the decrease in $L_{A,\rm ph}/(4\pi r_\star^2)$ causes cooler $T_{\rm co}$, faster $v_{\rm wind}$ and smaller $\dot{M}/(4\pi r_\star^2)$.
      It is beyond the scope of present study how these energy conversion efficiencies are determined, but we mentioned that the energy conversion efficiency from $L_{A,\rm co}$ to $L_{\rm kin,wind}$ ($\alpha_{{\rm wind}/A}$) is independent of stars, chromospheric magnetic field strengths, and energy inputs from the photosphere (Section \ref{sec:tco_empirical}). $\alpha_{{\rm wind}/A}$ is possibly dependent on the filling factor of open flux tube (Appendix \ref{sec:appendix_fph}).}
    \label{fig:figure19.eps}
  \end{center}
\end{figure*}

  Figure \ref{fig:figure19.eps} summarizes the causal relations among the varying $v_{\rm wind}$, $T_{\rm co}$, $\dot{M}$, $p_{\rm tr}$ and $\rho_{\rm co}$ with respect to decreases in $L_{A,\rm ph}/(4\pi r_\star^2)$ and $r_\star(T_{\rm eff})$. Here, $L_{A,\rm ph}/(4\pi r_\star^2)$ and $r_\star(T_{\rm eff})$ correspond to the vertical and horizontal axes of panels in Figure \ref{fig:figure18.eps}, respectively. As depicted with the thick arrows in Figure \ref{fig:figure19.eps}, the decrease in $L_{A,\rm ph}/(4\pi r_\star^2)$ causes cooler $T_{\rm co}$, faster $v_{\rm wind}$ and smaller $\dot{M}/(4\pi r_\star^2)$. In more detail, the smaller $L_{A,\rm ph}/(4\pi r_\star^2)$ is, the smaller $|L_{c,\rm co}|/(4\pi r_\star^2)$ and $L_{\rm kin,wind}/(4\pi r_\star^2)$ are. This energy partitioning from $L_{A,\rm ph}$ to $L_{c,\rm co}$ or $L_{\rm kin,wind}$ is represented by $\alpha_{{\rm wind}/A}$ (Section \ref{sec:tco_empirical} and Appendix \ref{sec:appendix_fph}).

  Smaller $|L_{c,\rm co}|$ tends to be associated with cooler $T_{\rm co}$ (Equation (\ref{eq:fcco_vs_tco})), which leads to lower $\beta_{\phi\max}$ (Equation (\ref{eq:beta_scale})). $v_{\phi\max}$ is amplified when $\beta_{\phi\max}$ is lower (Equation (\ref{eq:eq_vphimax_beta})), which is responsible for greater $(\Delta_p+\Delta_g)$ and faster $v_{\rm wind}$ (Section \ref{sec:vwind_empirical}). Smaller $L_{\rm kin,wind}/(4\pi r_\star^2)$ and faster $v_{\rm wind}$ drive smaller $\dot{M}/(4\pi r_\star^2)$ (Equation (\ref{eq:mdot_la_v^2})). On the other hand, smaller $r_\star$ (or cooler $T_{\rm eff}$) is associated with smaller $l_{B,\rm co}$ and $l_{B,\rm tr}$. The former leads to cooler $T_{\rm co}$ as well as smaller $L_{A,\rm ph}/(4\pi r_\star^2)$ (Equation (\ref{eq:fcco_vs_tco})). The effects of smaller $l_{B,\rm tr}$ and cooler $T_{\rm co}$ on $p_{\rm tr}$ are cancelled out with each other (Equation (\ref{eq:eq_p_tr_T_cr}), but smaller $l_{B,\rm tr}$ tends to enhance $p_{\rm tr}$ in general. As a result, cooler $T_{\rm co}$ and larger $p_{\rm tr}$ lead to larger $\rho_{\rm co}$.

%

\section{Discussion}
\subsection{Mass-Loss Rate of M-dwarf's Stellar Wind}
\label{sec:discussion_massloss}

\begin{figure*}
  \begin{center}
    \epsscale{1.}
    \plotone{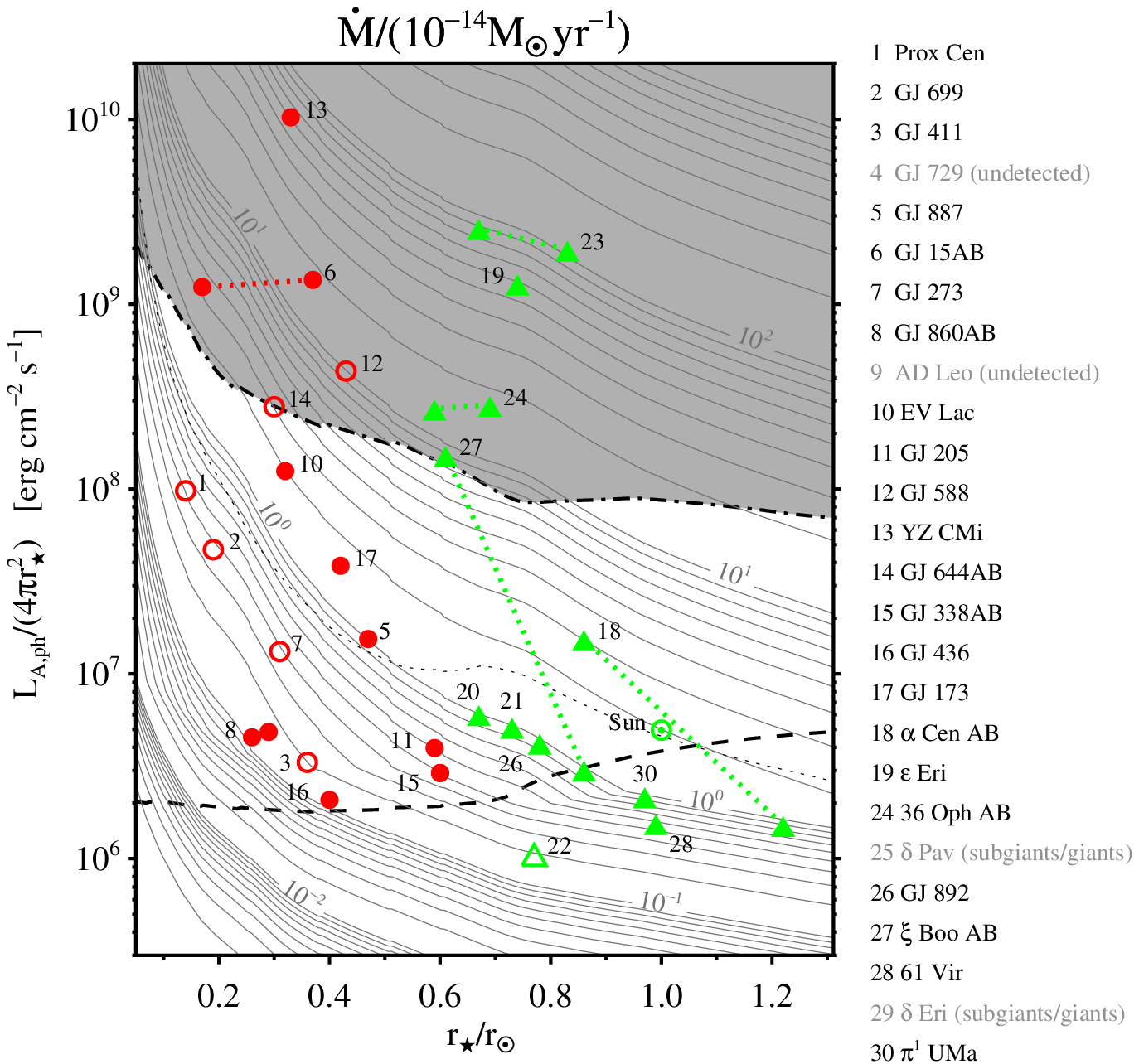}
    \caption{$r_\star$-$L_{A,\rm ph}$ diagram for mass loss rate ($\dot{M}$) with the observed $\dot{M}$ of main-sequence stars \citep{2017MNRAS.470.4026V,2021arXiv210500019W}.
        The red circle and green triangle symbols correspond to the positions of $(r_\star,L_{A,\rm ph})$ to reproduce the observed $\dot{M}$ of M and G, K dwarfs, respectively, where $\dot{M}$ of star indicated by open symbol is constrained only by an upper limit.
        The red and green dotted lines indicate the pairs of binaries.
        The numbers assigned to the symbols are ID numbers of stars used in \cite{2021arXiv210500019W} which are listed in the right side of figure. The star names in gray are not plotted in figure because they are not main-sequence stars or $\dot{M}$ are not detected.
      The black thick dashed and dash-dotted lines in $r_\star$-$L_{A,\rm ph}$ diagram mean the same as those in Figure \ref{fig:figure18.eps}. Thus, $L_{A,\rm ph}$ in the gray area is impossible to be generated even if the velocity of convective motion reaches the sound speed of stellar photosphere.}
    \label{fig:figure20.eps}
  \end{center}
\end{figure*}

The simulated mass-loss rates ($\dot{M}$) of M-dwarfs' stellar winds are much smaller than the solar wind's value (Table \ref{table:table_masslossrate}). This is the result mainly of the smaller surface areas of M dwarfs and their relatively faster winds. In addition, our wind's mass-loss rates of M dwarfs are typically smaller than reported by the previous global stellar wind modelings. $\dot{M}$ of M8 type star in this study is no more than $6.9\times10^{-17}$ $M_\odot$ yr$^{-1}$ while \cite{2017ApJ...843L..33G} and \cite{2018PNAS..115..260D} show $3\times10^{-14}$ $M_\odot$ yr$^{-1}$ and $4.1\times10^{-15}$ $M_\odot$ yr$^{-1}$ for TRAPPIST-1 (M8), respectively. $\dot{M}$ of Proxima Centauri (M5.5) by \cite{2016ApJ...833L...4G} and EV Lac (M3.5) by \cite{2014ApJ...790...57C} are $1.5\times10^{-14}$ $M_\odot$ yr$^{-1}$ and $3\times10^{-14}$ $M_\odot$ yr$^{-1}$, respectively, which are $10-100$ times higher than reproduced in our simulation. Because the inner boundary condition is arbitrarily determined in these 3D simulations \citep{2013ApJ...764...23S,2014ApJ...782...81V},  Alfv\'en wave energy flux on their inner boundary is likely to be inappropriately high for M dwarfs. Furthermore, because the spatial resolution of these modelings is too low to discuss the crucial role of compressible waves (slow shock) in the stellar wind acceleration, it is probable that they may not reproduce our simulation result.

Observational measurements of stellar wind have recently made remarkable progress especially in M dwarfs. In order to quantify the stellar wind's properties observationally, \citet{2001ApJ...547L..49W,2002ApJ...574..412W,2005ApJ...628L.143W,2005ApJS..159..118W,2021arXiv210500019W} investigated the absorption signatures in stellar Ly$\alpha$ spectra which originates in the ``neutral hydrogen wall'' around the astrospheres.
  By assuming that stellar winds' velocity are constant regardless of the stars, they succeeded in estimating mass-loss rates ($\dot{M}$) of several nearby stars.
  Another method to estimate $\dot{M}$ has been developed by \cite{2016A&A...591A.121B}, \cite{2017MNRAS.470.4026V} and \cite{2021MNRAS.501.4383V}, who deduced $\dot{M}$ of GJ 436 (M2.5) around $(0.45-2.5)\times10^{-15}$ $M_\odot$ yr$^{-1}$ by analyzing the transmission spectra of Ly$\alpha$ of GJ 436 b (a warm Neptune).

  The number of observations of M-dwarf's $\dot{M}$ drastically increases owing to \cite{2021arXiv210500019W}. The published $\dot{M}$ of M and G, K dwarfs are plotted in $r_\star$-$L_{A,\rm ph}$ diagram for $\dot{M}$ (Figure \ref{fig:figure20.eps}), which is obtained by converting $T_{\rm eff}$-$L_{A,\rm ph}$ diagram for $\dot{M}$ in Figure \ref{fig:figure18.eps} with the relation between $r_\star$ and $T_{\rm eff}$ for the main-sequence stars (Appendix \ref{sec:MS_r_M}).

  Figure \ref{fig:figure20.eps} shows the required Alfv\'en wave energy flux on the stellar photosphere ($L_{A,\rm ph}/(4\pi r_\star^2)$) to realize the observed $\dot{M}$. The red filled circle of GJ 436, for example, lies in the vicinity of the black thick dashed line, indicating that the Alfv\'en wave energy flux driven by the surface convective motion is almost adequate to reproduce the observed $\dot{M}$ of GJ 436. On the other hand, the observed $\dot{M}$ of EV Lac or YZ CMi could require nearly 50 or 5,000 times larger Alfv\'en wave energy flux than expected from the scenario of Alfv\'en wave generation by surface convective motion. In particular, YZ CMi lies in the gray area, which means that the Alfv\'en-wave driven stellar wind model discussed here cannot account for the observed $\dot{M}$ even if the velocity of convective motion exceeds the sound speed of the stellar photosphere. There are many data points suggesting much larger Alfv\'en wave energy flux is required than expected, similarly to EV Lac and YZ CMi. This discrepancy between the observed and predicted $\dot{M}$ might be resolved by investigating the contribution of coronal mass ejection to the total mass-loss rate, or the origin of Alfv\'en wave energy flux on the stellar surface.\\

\begin{figure*}
  \begin{center}
    \epsscale{1.}
    \plotone{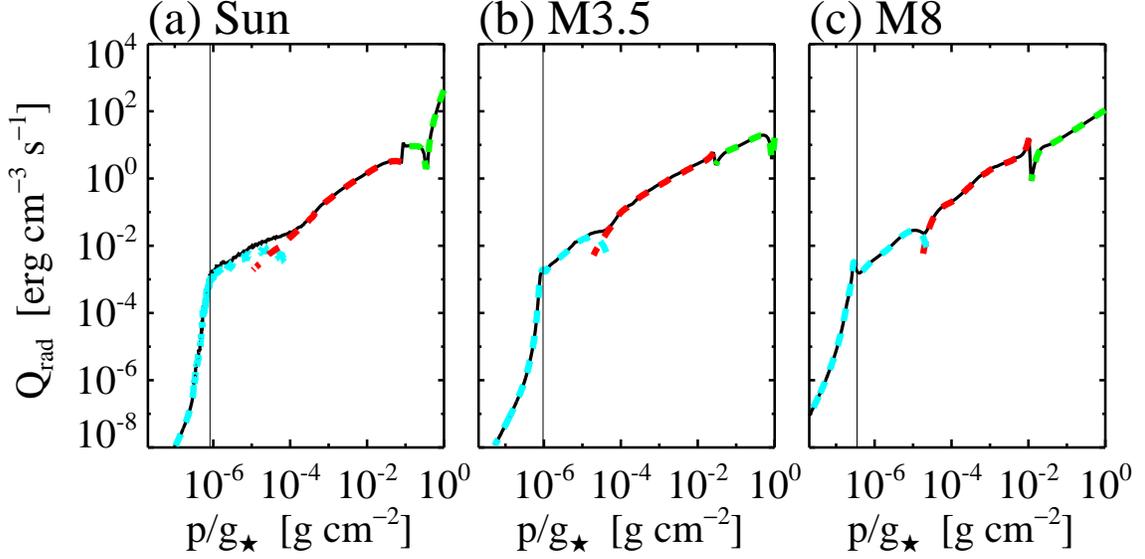}
    \caption{The temporally averaged radiative cooling of stellar atmospheres as a function of $p/g_\star$, representing the column mass density (black lines). The green, red, blue lines correspond to the cooling rates due to the photospheric, chromospheric, and coronal radiations, respectively ($Q_{\rm ph}$, $Q_{\rm ch}$, $Q_{\rm co}$ in Section \ref{sec:Heat Conduction and Radiative Cooling}). The vertical thin line shows the position of transition region (i.e., $p_{\rm tr}/g_\star$).}
    \label{fig:figure21.eps}
  \end{center}
\end{figure*}

\begin{figure*}
  \begin{center}
    \epsscale{1.}
    \plotone{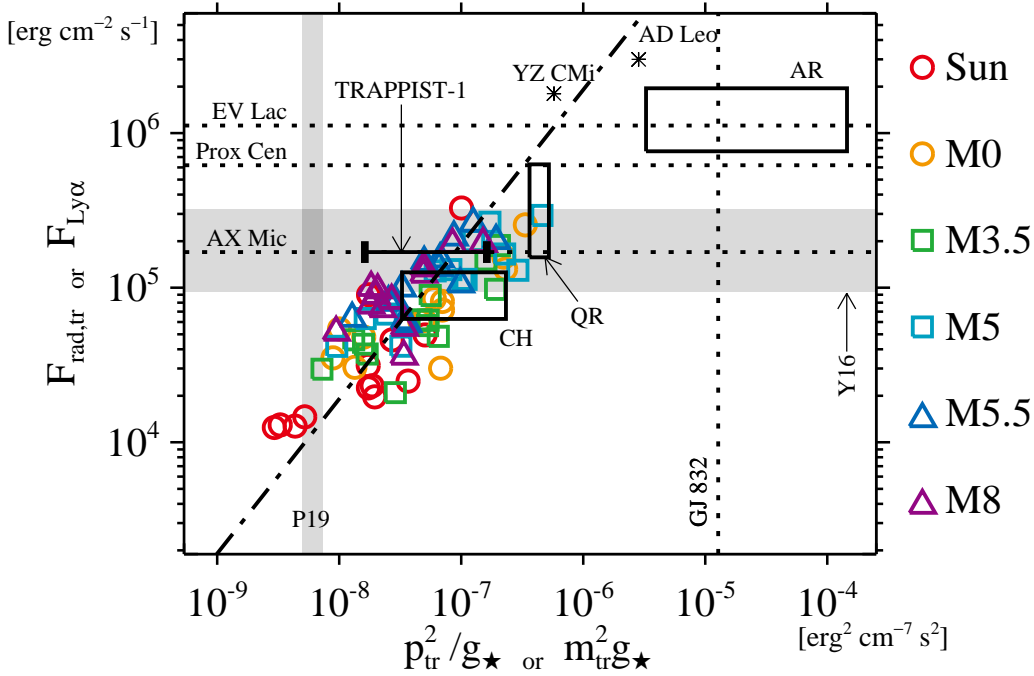}
    \caption{The properties of M dwarfs' transition region in our simulation. The vertical axis shows the radiative losses from the transition region in our simulation and the observed stellar Ly$\alpha$ fluxes. The horizontal axis shows the quantity of $p_{\rm tr}^2/g_\star$ where $p_{\rm tr}$ is the plasma pressure at transition region from our simulation or other literatures. $F_{\rm Ly\alpha}$ of M dwarfs are reported by \cite{1989A&A...208..159B} (for YZ CMi and AX Mic), \cite{2005ApJS..159..118W} (for AD Leo, EV Lac, and Proxima Centauri), \cite{2016ApJ...824..101Y} (for GJ 176, GJ 436, GJ 581, GJ 667C, GJ 832, GJ 876, and GJ 1214; they are indicated together by ``Y16'') and \cite{2017A&A...599L...3B} (for TRAPPIST-1).
      $F_{\rm Ly\alpha}$ from the solar active region (AR), quiet region (QR), and coronal hole (CH) is estimated based on \cite{1988ApJ...329..464F,1999ApJ...518..480F,2008A&A...492L...9C,2009ApJ...703L.152T}.
      $p_{\rm tr}$ of M dwarfs are reported by \cite{1982ApJ...258..740G} (for YZ CMi), \cite{1994A&A...281..129M} (for AD Leo), \cite{2016ApJ...830..154F} (for GJ 832), \cite{2019ApJ...871..235P} (for TRAPPIST-1), \cite{2019ApJ...886...77P} (GJ 176, GJ 436 and GJ 832; they are indicated by ``P19''. $p_{\rm tr}$ in solar coronal hole (CH) and quiet region (QR) are cited from \cite{1977ApJ...215..919M}, while $p_{\rm tr}$ in solar active region (AR) are cited from \cite{2001ApJ...550L.113Y}.
      The dash-dotted line represents a fitted line obtained by assuming the proportional relation between $p_{\rm tr}^2/g_\star$ and $F_{\rm rad,tr}$ ($y=1.90\times10^{12}x$).}
    \label{fig:figure22.eps}
  \end{center}
\end{figure*}

\begin{figure*}
  \begin{center}
    \epsscale{1.}
    \plotone{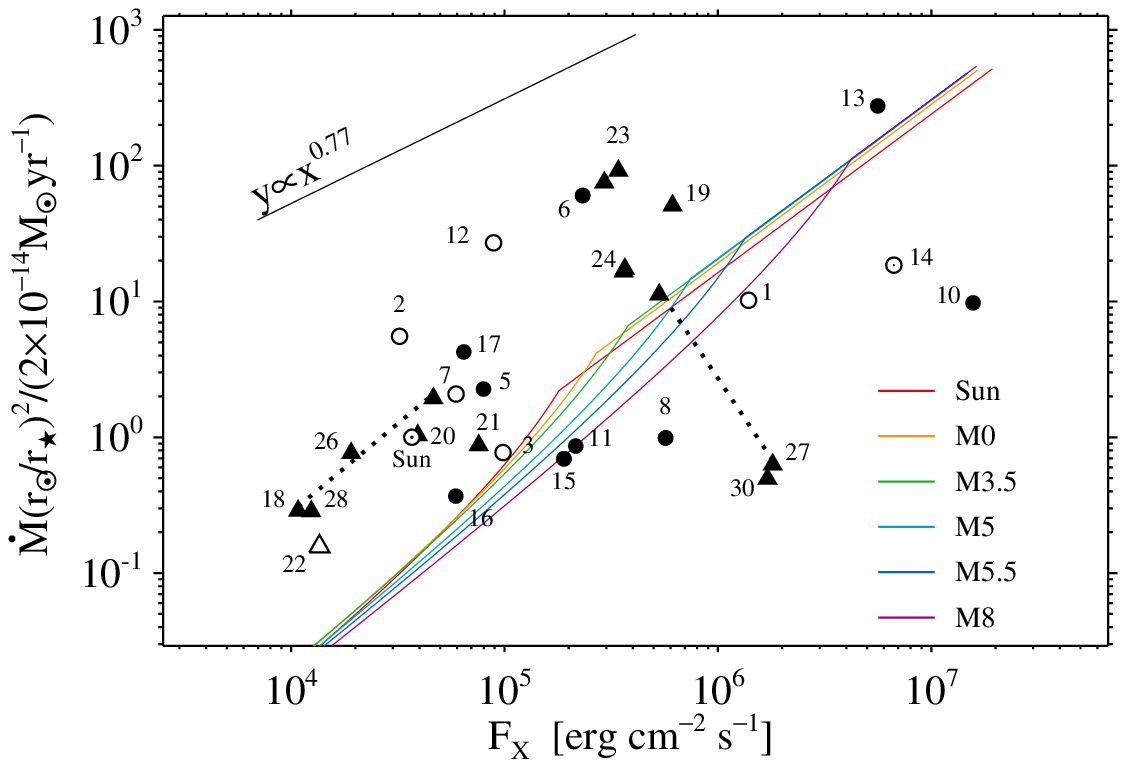}
    \caption{
        The predicted relation between X-ray flux ($F_X$) and wind's mass-loss rate ($\dot{M}$) based on this study (colored curves). The circle and triangle symbols correspond to the observed ($F_X$, $\dot{M}$) of M and G, K dwarfs, respectively, where $\dot{M}$ of star indicated by open symbol is constrained only by an upper limit.
        The dotted lines indicate the pairs of binaries.
        The numbers assigned to the symbols are ID numbers of stars used in \cite{2021arXiv210500019W}, and the corresponding star names are listed in Figure \ref{fig:figure20.eps}.
        $y\propto x^{0.77}$ is a fitted line derived by \cite{2021arXiv210500019W}.
    }
    \label{fig:figure23.eps}
  \end{center}
\end{figure*}

\subsection{Properties of Stellar Transition Region}
Although the detailed prediction of stellar spectrum from the simulated stellar atmosphere is beyond our scope, some implications are obtained to be compared with the observational studies. One of them is the plasma pressure at the stellar transition region $p_{\rm tr}$, which has been investigated by spectral analyses of optical lines or ultraviolet lines as one of the fundamental parameters to constrain the semi-empirical model atmosphere \citep{2017ARA&A..55..159L}.
Another is the radiative loss from the optically thin plasma around the transition region $F_{\rm rad,tr}$, which we define by using $Q_{\rm rad}$ in our simulation as follows:
\begin{equation}
  F_{\rm rad,tr}={1\over 4\pi r_\star^2}\int_{\rm tr}^\infty4\pi r^2 Q_{\rm rad}dr
  \label{eq:def_fradtr}
\end{equation}
where we assume that the transition region's emission is uniform over the whole stellar disk. Figure \ref{fig:figure21.eps} shows $Q_{\rm rad}$ of solar and stellar atmospheres as a function of $p/g_\star$.\par

Figure \ref{fig:figure22.eps} shows the relation between $p_{\rm tr}^2/g_\star$ and $F_{\rm rad,tr}$ obtained from our simulation. They are expected to be proportional to each other when $Q_{\rm rad}/p^2(\approx\Lambda(T)/(2k_BT)^2)$ steeply drop from the transition region to the corona. In Figure \ref{fig:figure22.eps}, we also show the observed stellar Ly$\alpha$ flux ($F_{\rm Ly\alpha}$), and $m_{\rm tr}g_\star$ which is obtained from the spectral analyses or emission measure diagnostics. Ly$\alpha$ line is formed from the upper chromosphere to the transition region \citep{2016ApJ...830..154F}, and generally the brightest and dominant emission line in the far-ultraviolet spectra of late-type stars \citep{1993ApJ...408..305L,2013ApJ...763..149F}. According to \cite{1995A&A...294..773H}, Ly$\alpha$ fluxes of M dwarfs are almost proportional to the column mass density at transition region. $F_{\rm Ly\alpha}$ of M dwarfs in Figure \ref{fig:figure22.eps} are reported by \cite{1989A&A...208..159B} (for YZ CMi and AX Mic), \cite{2005ApJS..159..118W} (for AD Leo, EV Lac, and Proxima Centauri), \cite{2016ApJ...824..101Y} (for GJ 176, GJ 436, GJ 581, GJ 667C, GJ 832, GJ 876, and GJ 1214; they are indicated together by ``Y16'' in Figure \ref{fig:figure22.eps}) and \cite{2017A&A...599L...3B} (for TRAPPIST-1). Note that M dwarfs reported by \cite{2016ApJ...824..101Y} are the targets of {\it Measurements of the Ultraviolet Spectral Characteristics of Low-mass Exoplanetary Systems} (MUSCLES) Treasury Survey \citep{2016ApJ...820...89F}, which are optically inactive and known to be planet-hosting stars.
We also show $F_{\rm Ly\alpha}$ from the solar active region (AR), quiet region (QR), and coronal hole (CH) \citep{1988ApJ...329..464F,1999ApJ...518..480F,2008A&A...492L...9C,2009ApJ...703L.152T}. Note that we converted the observed intensity of Ly$\alpha$ ($I_{\rm Ly\alpha}$) to $F_{\rm Ly\alpha}$ by assuming $F_{\rm Ly\alpha}=\pi I_{\rm Ly\alpha}$. The observed $F_{\rm Ly\alpha}$ are much larger than $F_{\rm rad,tr}$ for flare stars such as AD Leo, as expected because these stars are believed to be largely covered by active regions \citep{1982ApJ...260..670L,1985ApJ...299L..47S}. On the other hand, $F_{\rm Ly\alpha}$ are comparable to $F_{\rm rad,tr}$ for moderately active stars such as TRAPPIST-1 and MUSCLES targets. 

The plasma pressures at transition region ($p_{\rm tr}$) in Figure \ref{fig:figure22.eps} are reported by \cite{1982ApJ...258..740G} (for YZ CMi), \cite{1994A&A...281..129M} (for AD Leo), \cite{2016ApJ...830..154F} (for GJ 832), \cite{2019ApJ...871..235P} (for TRAPPIST-1), \cite{2019ApJ...886...77P} (GJ 176, GJ 436 and GJ 832; they are included in MUSCLES targets and indicated by ``P19'' in Figure \ref{fig:figure22.eps}). $p_{\rm tr}$ in solar coronal hole (CH) and quiet region (QR) are cited from \cite{1977ApJ...215..919M}, while $p_{\rm tr}$ in solar active region (AR) are cited from \cite{2001ApJ...550L.113Y}.

The simulated values of $p_{\rm tr}$ are significantly lower than that of flare stars but comparable to that of moderately active stars, similarly to the trend seen in the comparison between $F_{\rm rad,tr}$ and $F_{\rm Ly\alpha}$. On the other hand, it should be noted that the plasma pressure at stellar transition region is sometimes not well constrained from the semi-empirical modeling of stellar atmosphere. Indeed, the plasma pressure at transition region of GJ 832 reported by \cite{2019ApJ...886...77P} is two orders of magnitude lower than that by \cite{2016ApJ...830..154F}. The atmospheric modeling from the spectral analysis possibly depends on the model of micro-turbulent velocity \citep{2000A&A...358..575J}, partial frequency redistribution (PRD) effect, and ionization mechanisms due to the coronal back-heating. Our model of M dwarf's chromosphere and transition region also ignores the detailed radiative transfer and partial ionization effects, and therefore, much more effort is required to develop the realistic atmosphere model for M dwarfs.

  We finally present the predicted relation between X-ray flux ($F_X$) and wind's mass-loss rate ($\dot{M}$) based on this study (colored curves in Figure \ref{fig:figure23.eps}). The observed relation between them are provided by \cite{2021arXiv210500019W} (symbols in Figure \ref{fig:figure23.eps}). To derive these prediction curves, we assume that $F_X$ is represented by $F_{\rm rad,tr}$ (Equation (\ref{eq:def_fradtr})) and that $F_{\rm rad,tr}=1.90\times10^{12}p_{\rm tr}^2/g_\star$ (Figure \ref{fig:figure22.eps}). Because Figure \ref{fig:figure18.eps} shows that both $\dot{M}$ and $p_{\rm tr}$ are expressed as the functions of $(T_{\rm eff},L_{A,\rm ph})$, the relation between $F_{\rm rad,tr}$ and $\dot{M}$ is obtained for each $T_{\rm eff}$ with $L_{A,\rm ph}$ as an auxiliary variable.\\
  \indent The prediction curves in Figure \ref{fig:figure23.eps} suggest the tight correlation between $F_{\rm rad,tr}$ and $\dot{M}$, around which the data points scatter. However, regarding to the correlation between them, it is often pointed out that X-ray radiation from an active star originates in the plasma confined in the closed coronal loops, and that, even in the case of the Sun, X-ray flux varies by an order of magnitude during its activity cycle \citep{2011MNRAS.417.2592C}.

\section{Summary}
We summarize the conclusions about the differences and similarities in the stellar chromospheres, coronae, and winds among the Sun and M dwarfs. These findings are obtained by analyzing the results of parameter survey based on the one-dimensional magnetohydrodynamics numerical simulations.\\
\noindent
(i) Regardless of the Sun or M dwarfs, the nonlinear propagation of Alfv\'en wave is responsible for driving the stellar spicule, heating the stellar atmosphere, and accelerating the stellar wind (Section \ref{sec:results}).\\
\indent
The following (ii), (iii), (iv) are the arguments for a given transmitted Alfv\'en wave energy flux into the corona ($F_{A,\rm co}$).\\
(ii) M dwarf's corona tends to be cooler and denser than solar corona (Section \ref{sec:coronal_temperature}, \ref{sec:p_tr}). The shorter spatial scale of coronal magnetic field of M dwarf results in this tendency. 
\\
(iii) M dwarfs' stellar winds are relatively faster than the solar wind (Section \ref{sec:vwind}). The lower plasma $\beta$ of M-dwarf's stellar wind is a suitable environment for Alfv\'en wave amplification, which leads to the generation of stronger slow shock and further acceleration of stellar wind (Section \ref{sec:vwind_empirical}).\\
(iv) The mass-loss rates of M dwarfs' stellar winds are much smaller than that of solar wind, because of (1) the much smaller stellar surface area, (2) the constant energy conversion efficiency from the Alfv\'en wave energy flux in the corona to the kinetic energy flux of the stellar wind ($\alpha_{{\rm wind}/A}$ in Section \ref{sec:tco_empirical}), and (3) the relatively faster stellar wind.\\
(v) The transmissivity of Alfv\'en wave energy flux from the photosphere to the corona ($\alpha_{\rm co/ph}$) is determined mainly by the Alfv\'en travel time between them ($\tau_{A,\rm co}$ in Section \ref{sec:transmissivity}). $\alpha_{\rm co/ph}$ is negatively correlated with $\tau_{A,\rm co}$ when $\tau_{A,\rm co}\gtrsim\nu_{\rm ac}^{-1}$ ($\nu_{\rm ac}$ is the acoustic cutoff frequency), while the correlation is reversed when $\tau_{A,\rm co}\lesssim\nu_{\rm ac}^{-1}$ (Figure \ref{fig:figure16.eps}).
The stronger chromospheric magnetic field ($\overline{B}$) reduces $\tau_{A,\rm co}$ (Equation \ref{eq:tau_A_nu_ac_scale}) and affects the transmissivity of Alfv\'en wave.\\
(vi) We developed the semi-empirical formulae to estimate $v_{\rm wind}$, $T_{\rm co}$, $\dot{M}$, $p_{\rm tr}$, $\rho_{\rm co}$ from a given combination of $L_{A,\rm ph}$ and $T_{\rm eff}$ (Figure \ref{fig:figure18.eps}). Comparison of them to the observations suggest that the observed $\dot{M}$ of EV Lac or YZ CMi could require nearly 50 or 5,000 times larger Alfv\'en wave energy flux than expected from the scenario of Alfv\'en wave generation by surface convective motion (Section \ref{sec:discussion_massloss}).

  \section{Future Works}
  The proposed semi-empirical formulae would contribute to the studies on the dynamics of planetary atmosphere or magnetosphere \citep{2007AsBio...7..167K,2007AsBio...7..185L,2019LNP...955.....L}, the evolution of stellar rotation \citep{2017ApJ...849...83P,2020ApJ...896..123S}, or the astrospheres' structures of main-sequence stars \citep{2021arXiv210500019W}. It would be helpful as an initial guess to constrain the physical quantities of stellar chromosphere, corona and wind. However, there remain so many unresolved problems in this paper. That means further development of simulation scheme and sophisticated interpretation are required to comprehend the physics of stellar atmosphere and wind in more unified way.\\
  \indent
  The present numerical scheme is based on a 1D approximation (axisymmetry assumption), by which the global magnetic field configuration and its interaction with stellar wind flow cannot be addressed.
  It is likely that the other simplifications in this study limit the applicability of our results. The partial ionization effects and collisionless effects are involved with the dissipation of Alfv\'en wave in the stellar chromosphere and interplanetary space, respectively, but not considered in this study. Alfv\'en wave turbulence is also important process for heating stellar atmosphere and wind. \\
  \indent
  As for our interpretation of simulation results, it should be kept in mind that there are many heuristic relations adopted in our semi-empirical formulae; e.g., the simplified energy conservation law ($L_{A,\rm co}\approx L_{\rm kin,wind}-L_{g,\rm co}-L_{c,\rm co}$; Equation (\ref{eq:energy_conservation_app})), the constant energy conversion efficiency ($\alpha_{{\rm wind}/A}=L_{\rm kin,wind}/L_{A,\rm co}=$const.), or the relation between the maximum nonlinearity of Alfv\'en wave in the stellar wind and the plasma $\beta$ of stellar wind (Equation (\ref{eq:eq_nonlinearity_beta})). The efforts for validation or generalization of these assumptions will lead to more unified understanding of stellar atmosphere and wind physics from the Sun to M dwarfs.

\acknowledgments
T.S. was supported by JSPS KAKENHI Grant Number JP18J12677. K.S. was also supported by JSPS KAKENHI Grant Number 21H01131. A part of this study was carried out by using the computational resources of the Center for Integrated Data Science, Institute for Space-Earth Environmental Research, Nagoya University through the joint research program, XC40 at YITP in Kyoto University, and Cray XC50 at Center for Computational Astrophysics, National Astronomical Observatory of Japan. Numerical analyses were partly carried out on analysis servers at the Center for Computational Astrophysics, National Astronomical Observatory of Japan.

\appendix

\section{Effect of varying filling factor of open flux tube}
\label{sec:appendix_fph}

While we discuss the simulation results with the fixed $f_{\rm ph}$ at $1/1600$, where $f_{\rm ph}$ is the filling factor of open magnetic flux tube, there are many observational and theoretical studies which point out the crucial role of $f_{\rm ph}$ in determining the solar wind velocity ($v_{\rm wind}$), especially the positive correlation between $f_{\rm ph}$ and $v_{\rm wind}$ \citep{1990ApJ...355..726W,2000JGR...10510465A,2006JGRA..111.6101S,2007ApJS..171..520C,2017SoPh..292...41T}. 

Moreover, $f_{\rm ph}$ of stars could generally vary within a wide range, depending on the rotation period, convective turnover time, Rossby number, amplitude of magnetic activity cycle, and so on. Even in the case of the Sun, $f_{\rm ph}$ can vary by an order of magnitude \citep{2000GeoRL..27..505W,2011ApJ...741...54C}. Although the filling factor of magnetically active region on the stellar surface has been investiageted by spectroscopic observations, it is still challenging to distinguish the filling factor associated with open flux tubes from the total filling factor \citep{2019ApJ...876..118S}.

To check the applicability of our semi-empirical method, thus, we performed the additional parameter survey about the stellar atmosphere and wind of M3.5 dwarf, by varying $f_{\rm ph}$ from 1/32 to 1/400, 1/1600, 1/6400, 1/32000, 1/64000, and 1/80000. The other parameters such as the chromospheric magnetic field strength ($\overline{B}$) and Alfv\'en wave amplitude of the photosphere ($v_{\rm ph}$) are set to $\overline{B}=B_{\rm ph}e^{-5}$ and $v_{\rm ph}=v_{\rm conv}$, respectively.

\begin{figure}
  \begin{center}
    \epsscale{.6}
    \plotone{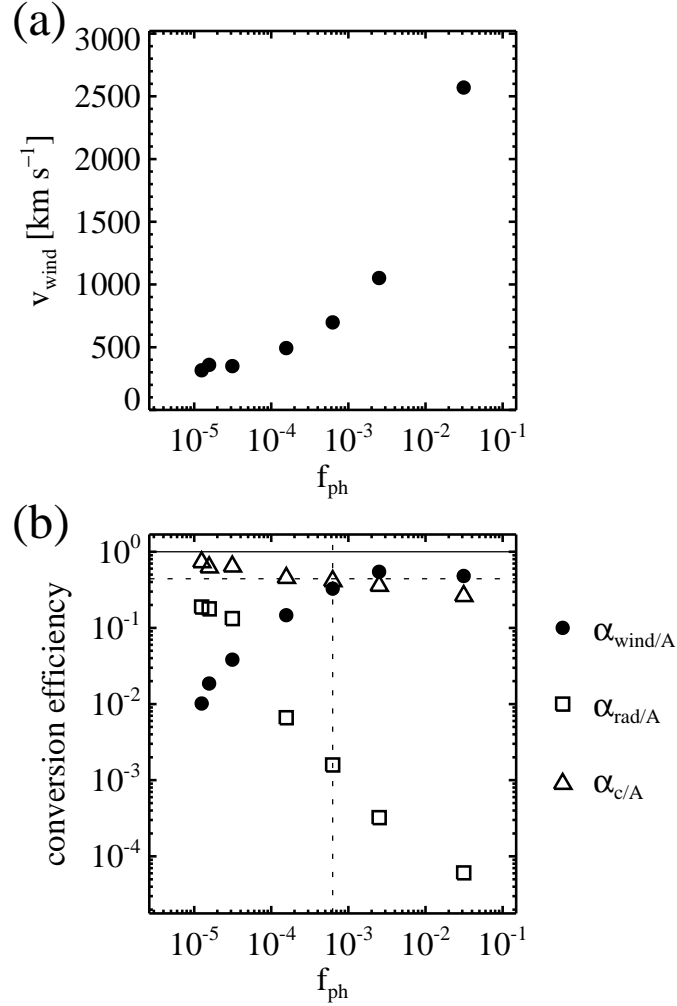}
    \caption{(a) The positive correlation between stellar wind velocity at $r=100r_\star$ ($v_{\rm wind}$) and filling factor of open flux tube ($f_{\rm ph}$). (b) $f_{\rm ph}$ vs energy conversion efficiencies from the transmitted Alfv\'en wave energy flux in the corona to (1) the kinetic energy of the stellar wind ($\alpha_{{\rm wind}/A}=L_{\rm kin,wind}/L_{A,\rm co}$; filled circles), (2) radiative energy loss in the corona ($\alpha_{{\rm rad}/A}=-L_{\rm rad,co}/L_{A,\rm co}$; open squares), and (3) heat conduction flux in the corona ($\alpha_{c/A}=-L_{c,\rm co}/L_{A,\rm co}$; open triangles). The vertical and horizontal dashed lines correspond to $f_{\rm ph}=1/1600$ and $\alpha_{{\rm wind/A}}=0.442$, respectively.}
    \label{fig:figure24.eps}
  \end{center}
\end{figure}

Figure \ref{fig:figure24.eps}(a) shows that the stellar wind velocity at $r=100r_\star$ ($v_{\rm wind}$) as a function of $f_{\rm ph}$. They are positively correlated with each other as found in the previous studies. Our semi-empirical method is, on the other hand, not useful to understand the physical mechanism leading to this relation. This is because the coefficients and power-law indices appearing in the semi-empirical formulae probably depend on $f_{\rm ph}$. One of the remarkable examples is the dependence of $\alpha_{{\rm wind}/A}$ on $f_{\rm ph}$, where $\alpha_{{\rm wind}/A}$ is the energy conversion efficiency from the transmitted Alfv\'en wave energy in the corona to the kinetic energy of the stellar wind; i.e., $\alpha_{{\rm wind}/A}=L_{\rm kin,wind}/L_{A,\rm co}$. The filled circles in Figure \ref{fig:figure24.eps}(b) shows $\alpha_{{\rm wind}/A}$ as a function of $f_{\rm ph}$. As shown in this panel, $\alpha_{{\rm wind}/A}$ steeply drops from $\sim0.4$ to $\sim 10^{-2}$ with decreasing $f_{\rm ph}$. This phenomenon is probably related to the rapid increase in the radiative energy loss with decreasing $f_{\rm ph}$. In Figure \ref{fig:figure24.eps}(b), the square symbols show the energy conversion efficiency from the transmitted Alfv\'en wave energy in the corona to the radiative energy loss; i.e., $\alpha_{{\rm rad}/A}=-L_{\rm rad,co}/L_{A,\rm co}$. Because flux tube with smaller $f_{\rm ph}$ is characterized with larger expansion ratio of $A/A_{\rm tr}$ ($A_{\rm tr}\leq A\leq A_{\rm co}$) and smaller $l_{B,\rm tr}$ (Equation (\ref{eq:l_b_tr_definition})), the plasma pressure at transition region (Equation (\ref{eq:eq_p_tr_final})) and coronal mass density (Equation (\ref{eq:eq_rho_cr_final})) increase for a given transmitted Alfv\'en wave energy flux into the corona, so that the radiative energy loss is enhanced. As a result, the approximated energy conservation law (Equation (\ref{eq:energy_conservation_app})) becomes invalid in the case of smaller $f_{\rm ph}$, which requires further improvement of our semi-empirical method to understand more comprehensively the mechanisms of heating stellar chromospheres, coronae and driving the stellar winds.

We note that the stellar atmosphere and wind structures would similarly depend on $H_l$; the parameter of typical closed loop height. Several radio observations suggest that the closed loop system of M dwarf extends to a few times stellar radii \citep{1995A&A...298..187B,2021arXiv210501021D}, while $H_l=0.1r_\star$ in this study. If we adopt $H_l={\rm a\ few}\times r_\star$ straightforwardly based on these observations, the stellar wind velocity will increase and mass-loss rate will decrease. This is because larger $H_l$ represents the stronger magnetic field in the distance and leads to shorter spatial scale of expanding flux tube ($l_{B,\rm co}$ in Equation (\ref{eq:l_b_definition})). The subsequent phenomena caused by shorter $l_{B,\rm co}$ are depicted in Figure \ref{fig:figure19.eps}. The stellar wind velocity ($v_{\rm wind}$) is further accelerated by more largely amplified Alfv\'en wave. Because $H_l$ does not so much affect the wind's kinetic energy luminosity ($L_{\rm kin,wind}$), faster $v_{\rm wind}$ results in smaller mass-loss rate.

\section{Main-sequence star's radius and mass}
\label{sec:MS_r_M}

To obtain Figure \ref{fig:figure18.eps}, we assume $T_{\rm eff}$-$r_\star$ and $r_\star$-$M_\star$ relations of main-sequence stars by referring to \citep{2012ApJ...746..101B,2019MNRAS.484.2674R}. The adopted relation of $T_{\rm eff}$-$r_\star$ is as follows.
\begin{align}
  {r_\star\over r_\odot}
  =\max&\left[-0.367+1.041\left(T_{\rm eff}\over T_{\rm eff\odot}\right),\right.\nonumber\\
    &\ \ -8.133+29.389\left(T_{\rm eff}\over T_{\rm eff\odot}\right)
    -32.8468\left(T_{\rm eff}\over T_{\rm eff\odot}\right)^2
    +12.4474\left(T_{\rm eff}\over T_{\rm eff\odot}\right)^3,\nonumber\\
    &\left.\ \ \left(T_{\rm eff}\over T_\odot\right)^{1.866}H(T_{\rm eff}-4000)\right],
\end{align}
where $H(x)$ is the Heaviside step function; i.e., $H(x)=0$ for $x<0$ and $H(x)=1$ for $x\geq0$. The obtained $r_\star(T_{\rm eff})$ is smoothed with a boxcar average of 500 K.\\
\indent
The relation of $r_\star$-$M_\star$ is defined as follows.
\begin{align}
  {M_\star\over M_\odot}=\max\left[x_M,\left(r_\star\over r_\odot\right)^{1.091}H(r_\star-0.5r_\odot)\right],
\end{align}
where $x_M$ is the solution of the following equation for a given $r_\star/r_\odot$.
\begin{align}
  {r_\star\over r_\odot}=0.013+1.238x_M-1.13x_M^2+1.21x_M^3.
\end{align}

  We used the $\AE$SOPUS opacity table published by \cite{2009A&A...508.1539M} to estimate the physical quantities such as mass density and plasma pressure on the photosphere for the stars with $T_{\rm eff}>4000$ K.


\bibliographystyle{aasjournal}
\bibliography{bibliography}



\end{document}